\documentclass[referee]{aa-mpa}
\usepackage{graphicx}
\makeatletter 
\let\@internalcite\cite
\def\cite{\@ifstar{\citeyear}{\citefull}}
\def\citefull{\def\astroncite##1##2{##1 ##2}\@internalcite}
\def\citeyear{\def\astroncite##1##2{##2}\@internalcite}
\def\citeau{\def\astroncite##1##2{##1}\@internalcite}
\def\citen{\def\astroncite##1##2{##1 (##2)}\@internalcite}
\def\possesivcite{\def\astroncite##1##2{##1's (##2)}\@internalcite}
\def\@citex[#1]#2{\if@filesw\immediate\write\@auxout{\string\citation{#2}}\fi
  \def\@citea{}\@cite{\@for\@citeb:=#2\do
    {\@citea\def\@citea{; }\@ifundefined
       {b@\@citeb}{{\bf ?}\@warning
       {Citation `\@citeb' on page \thepage \space undefined}}%
{\csname b@\@citeb\endcsname}}}{#1}}
\def\@cite#1#2{#1\if@tempswa , #2\fi}
\def\@biblabel#1{}
\makeatother
\begin{document}

\thesaurus{20(08.05.3; 08.09.3; 08.16.3)}

\title{Age--luminosity relations for low-mass metal-poor stars}

\author{A.~Weiss \and H.~Schlattl}

\institute{Max-Planck-Institut f\"ur Astrophysik,
           Karl-Schwarzschild-Str.~1, 85748 Garching,
           Federal Republic of Germany}

\offprints{A.~Weiss; (e-mail address: aweiss@mpa-garching.mpg.de)}
\mail{A.~Weiss}

\date{Received; accepted}

\authorrunning{Weiss \& Schlattl:}
\titlerunning{Age--luminosity relations for low-mass metal-poor stars}

\maketitle
\vspace{3.0cm}
\begin{abstract}
We present a grid of evolutionary calculations for metal-poor low-mass stars for
a variety of initial helium and metal abundances. The intention is mainly
to provide a database for deriving directly  stellar ages of halo and globular
cluster stars for which basic stellar parameters are known, but the tracks can
also be used for isochrone or luminosity function construction, since they
extend to the tip of the red giant branch. Fitting formulae for age-luminosity
relations are provided as well. The uncertainties of the evolutionary ages due
to inherent shortcomings in the models and due to the unclear effectiveness of
diffusion are discussed. A first application to field single stars is
presented. 
\end{abstract}
\clearpage

\section{Introduction}

The determination of stellar ages provides numerous clues on the
evolution of the Milky May and its components. While the classical
method for globular clusters relies on morphological features in the
Hertzsprung--Russell-diagram (HRD), for example an age-dependent
turn-off (TO) brightness, direct age determinations of individual stars is
progressively becoming a feasible alternative. In this case, the age
is obtained from the position of the star on an evolutionary
track. Typical examples for this method are eclipsing binary systems
such as AI~Phe (\cite{acg:88}; \cite{msk:92}) or $\zeta$~Aur
(\cite{bhb:96}; \cite{schpe:97}), for which accurate masses and radii can be
determined from the light-curves (photometry and spectroscopy needed),
and the stellar composition is either 
assumed or obtained from spectral analysis. In this case, the
requirement that both components should have the same age provides an
independent test for stellar evolution theory. If the distance to a
single star is known accurately (e.g.\ from HIPPARCOS parallaxes) and its
composition, gravity and effective temperature can be determined
spectroscopically, the same procedure can be applied as well. For example, this
method has recently been used for metal-poor 
stars by \citeau{fuhr:00} (\citeyear{fuhr:98}; \citeyear{fuhr:00}) to
deduce the formation history of bulge, 
thick and thin disk, and halo of the Milky Way. 
In the future, detached eclipsing binaries will hopefully be detected
in globular clusters by massive photometric searches (e.g.\ in
$\omega$~Cen, \cite{kks:96} and \cite{kks:97}, and in M4,
\cite{ktk:97}) and will then allow age determinations of individual
cluster stars (\cite{pacz:97}).

To be prepared for this case, we provide a grid of evolutionary tracks
for low-mass, metal-poor stars typical for Pop~II. The grid not only
extends over stellar mass (from $0.6\cdots 1.3 \,M_\odot$) but also
over composition, both in metallicity $Z$ and helium content $Y$. The
latter parameter usually is kept at a fixed value (typically 0.23) or
coupled to $Z$ via an assumption about the chemical evolution. Due to
the small absolute value of $Z$ for Pop~II stars, this is however an
almost negligible effect. According to \citen{pacz:97} it might be
feasible to determine age {\em and} helium content of members of
detached eclipsing binary systems and therefore calculations for
different values of the initial helium abundance are necessary. 

As a further effect microscopic diffusion has to be
considered. Although it has become evident that it is operating in the
Sun (\cite{rvc:96}; \cite{gd:97}), there are also arguments that
diffusion may not be as efficient as calculated, as can be seen from
the high abundance of $^7{\rm Li}$ still present in the
photosphere of metal-poor low-mass stars (\cite{vc:98}). Therefore, calculations with
and without diffusion have been performed to cover the whole possible
range.

In Sect.~2 we will summarize the main properties of our stellar
evolution code and the calculations done. After this, the results will
be presented. We make available tables with the full evolutionary
properties of all cases calculated. These can be used for isochrone
calculations as well, if needed. We avoid transformations into observed
colours and magnitudes for several reasons: firstly, such
transformations always involve a further source of uncertainty (see
\cite{ws:99} for a discussion); secondly, they can, if needed, easily be
applied, since the tables contain all necessary data; and finally, the
data expected from the analysis of binary lightcurves and from
spectroscopy will yield physical quantities, anyhow.  To facilitate the
derivation of the stellar age from given values of the global stellar
properties, fitting formulae and the corresponding coefficients will be
supplied as well. In Sect.~4 we will finally discuss the accuracy of
such direct age determinations with special emphasis laid upon the
comparison with other, independent work, because this will provide
insight into the inherent systematic uncertainties of theoretical
stellar evolution calculations. In the absence of suitable binary
systems, we have applied our results to a few nearby single stars with
known absolute parameters. This will be presented in Sect.~4 as well,
before the conclusions close the paper.

\section{The stellar evolution code}

We are using the Garching stellar evolution code, which is a
derivative of the original Kippenhahn-code (\cite{kwh:67}) developed
and improved over the years. While new properties of the code have
always been documented in the corresponding publications, we summarize
them here again for completeness:

\paragraph{Numerical aspects:} The Lagrangian spatial grid (in relative
mass $M_r/M$) adopts itself to structure changes. Its resolution is
controlled by an algorithm ensuring that the partial differential
equations are solved with a given accuracy (\cite{ww:94a}). 
Since all composition changes are calculated between two models of two
successive evolutionary ages, the evolution of temperature and density
during this time-step has to be given at each grid-point. We use a
predictor--corrector scheme for this (\cite{schl:96}; \cite{swl:97}). The
assumption of constant $T$ and $\rho$ can be used as an alternative,
but requires time-steps smaller by about a factor of 2--5. 
Nuclear burning and particle transport processes (convection and
diffusion) are calculated either simultaneously in a single iterative scheme
with a generalized Henyey-solver (\cite{schl:99}) or separately in a
burning--mixing--burning--\ldots sequence. In the latter case, the network
solves the linearized particle abundance equations in an implicit way.
In both cases a number of time-steps, which are smaller
than that between the two models and which are adopting to composition
changes, are followed until the whole evolutionary time-step is covered.

\paragraph{Opacities:} We use as the sources for the Rosseland mean
opacities the latest OPAL tables (\cite{ir:96}) and the molecular
opacities by \citen{af:94}. Both groups provided us (Rogers 1995,
private communication, and Alexander, 1995, private communication)
with tables for 
{\em exactly the same compositions} including the enhancement of
$\alpha$-elements (see \cite{sw:98} for details). The tables, which
have a common $T$-$\rho$-grid, can smoothly be merged and together
with electron conduction opacities (\cite{imi:83}) result in
consistent tables for all stellar interior conditions encountered. The
interpolation within a single table is done by bi-rational
two-dimensional splines (\cite{spaeth:73}), which contain a free
parameter allowing the transition from standard cubic to near-linear
interpolation. This avoids unwanted spline oscillations but guarantees
that the interpolant is always differentiable twice. We then
interpolate in a 3x3 cube in $X$--$Z$--space to the grid-point's
composition by two independent polynomial interpolations of degree
2. The cube of table compositions is chosen such that the central
point is closest to the actual composition under consideration.

\paragraph{Equation of state:} We use, where possible, the OPAL
equation of state (\cite{rsi:96}) with the interpolation procedures
provided by the same authors along with the EOS tables. For low
densities and temperatures, we use our traditional Saha-type EOS for a
partially ionized plasma or approximations for a degenerate electron
gas (see \cite{kwh:67}, \cite{weiss:87a}, \cite{wag:96} for
details). However, we did not calculate models not being covered by the
OPAL EOS for the larger part. This limits the mass range to  $\ge
0.6\,M_\odot$. 

\paragraph{Convection:} Convection is treated in the standard
mixing-length approach. No overshooting or semi-convection is
considered. The mixing-length parameter is calibrated with a solar
model calculated without diffusion; the resulting value for the
physical input employed here is 1.59 pressure scale heights. Note that it
would be slightly different for solar models including diffusion.
Convective mixing is either assumed to be instantaneous or treated as
a fast diffusive process in the case that all processes affecting the
chemical composition are treated simultaneously.

\paragraph{Neutrino emission:} Energy losses due to plasma processes are
included according to \citen{hrw:94} for plasma-neutrinos and
\citen{mki:85} for photo- and pair-neutrinos.

\paragraph{Nuclear reactions:} We use the \citen{cf:88} reaction rates
and the Salpeter formula for weak screening. The nuclear network
follows the evolution of $^1{\rm H}$, $^3{\rm He}$, $^4{\rm He}$, 
$^{12}{\rm C}$, $^{13}{\rm C}$, $^{14}{\rm N}$, $^{15}{\rm N}$, 
$^{16}{\rm O}$, $^{17}{\rm O}$. All other species in the
$pp$-chain and CNO-cycles are assumed to be in equilibrium (this is
justified because it assumes only that $\beta$-decays are faster than
$p$-captures). The network can also treat later burning
phases, but this is of no concern here because calculations were
stopped at the onset of the core helium flash.

\paragraph{Diffusion:} We consider the diffusion of hydrogen and helium. While
our code also allows for metal diffusion, we ignored this here for reasons of
CPU economy. Experience from solar models shows that the effect of
metal diffusion  on the interior evolution is only a fraction of that of
H/He diffusion. This is confirmed by test calculations in which metal
diffusion was included (see next section).
The various coefficients of the particle
diffusion equations are calculated according to \citen{tbl:94} with
the routine provided kindly by A.~Thoul (1997, private
communication). If diffusion is considered, it turned out to be both
more accurate and numerically stable to treat diffusion and burning in
a single numerical algorithm (see above).

\section{Calculations and results}

\subsection{Details of the calculations}

We have calculated two complete sets of models, one (canonical) without and one
with particle diffusion (denoted ``C'' and ``D''). In each set the
following values of mass and composition were explored:
\begin{enumerate}	
\item {\em Mass:} $M=0.6\cdots 1.3\,M_\odot$ in steps of $0.1\,M_\odot$ 
\item {\em Helium:} $Y=0.20,\; 0.25,\; 0.30$
\item {\em Metallicity:} $Z=0.0001,\; 0.0003,\; 0.001,\; 0.003$
\end{enumerate}
Due to the varying helium abundance,  $\rm [Fe/H]$ is not constant for
fixed $Z$. Table~\ref{t:fehv} lists $\rm [Fe/H]$ for all mixtures.

{\renewcommand\baselinestretch{1.0}
\begin{table}[ht]
\caption{$\rm [Fe/H]$ for the twelve initial compositions, for which
evolutionary tracks have been computed.}
\protect\label{t:fehv}	
\begin{tabular}{c|rrrr}
$\downarrow\,Y\,\vert\,Z\,\rightarrow$ & 0.0001 & 0.0003 & 0.0010 & 0.0030 \cr
\hline
0.20 & -2.645 & -2.168 & -1.644 & -1.166 \\
0.25 & -2.617 & -2.140 & -1.616 & -1.138 \\
0.30 & -2.587 & -2.110 & -1.586 & -1.108 \\
\end{tabular}
\end{table}
}

The metallicity range is that for typical globular clusters, but is not covering
the most metal-rich ones like 47~Tuc or M107. The same enhancement of
$\alpha$-elements (\cite{sw:98}) is always assumed and is the one for
which we have the opacity tables available. 
$Z$ therefore denotes the total metallicity including the $\alpha$-element
enhancement in all cases. 
Because $\alpha$-element enhancement is typical for Pop~II stars,
no calculation for solar metal ratios has been done. We recall that
for very low
metallicities, the evolution depends primarily on the total metallicity
(\cite{ssc:93}) and only slightly on the internal metal
distribution. This, however,
becomes non-negligible at the upper end of our metallicity range. For
example, \citen{sw:98} find that already at $Z=0.002$ the turn-off of
isochrones is about 0.05~mag brighter for models which include
$\alpha$-enhancment as compared to a solar-scaled mixture with
identical $Z$. Also, the RGB colour is bluer by $\approx$~0.05~mag. 

It is not yet clear whether the amount of oxygen enhancement in
metal-poor stars is independent of metallicity (see
\cite{gsc:00} for a recent result), or whether there are systematic
variations of  $\rm [O/Fe]$ with $\rm [Fe/H]$ ( see \cite{ilr:98} for
unevolved metal-poor stars). In both cases, the oxygen enhancement of
our metal mixture ($\rm [O/Fe]=0.5$) is a good representation of the
average enhancement in the metallicity range under consideration. The
same is true for the magnesium overabundance (\cite{fuhr:98}; see also
\cite{sw:98} for the spread of abundances of other
$\alpha$-elements). Variations around the mean $\alpha$-enhancement
are a second-order effect, which could be considered only in modeling
individual objects, provided the availability of appropriate opacity tables.

The helium values of our mixtures were chosen as to
certainly cover the possible range, with the central value of $Y=0.25$
being close to a primordial value of $0.244\pm0.002$
(\cite{it:98}). This value is somewhat higher than the traditionally
assumed 0.23, which, however is too low, even for the more generally
accepted primordial value of $0.234\pm0.002$ (\cite{oss:97}).
Different initial helium contents in the calculations allow save
interpolation to any prefered value or to keep it as a free
parameter. Some additional mixtures were considered for specific mass
values in order to be able to compare with published results (see
Sect.~4.1). 

All calculations were started from homogeneous zero-age main-sequence
(ZAMS) models with vanishing gravothermal energies. This implies
adjustment of isotopes to their equilibrium values in the stellar core;
this period lasts for several $10^7$ years. The resulting small loop in
the HRD is omitted in all figures and tables and the ages reset to zero
for the models with minimum gravothermal energy production. This
definition does not necessarily coincide with the minimum luminosity
during the initial loop, which would be an alternative choice for the
ZAMS position. For the lower masses our definition corresponds to ages
of a few $10^7$~yrs, for the higher masses to about $10^5$~yrs or even
less. The evolution is followed up to the tip of the red giant branch
(RGB), when helium violently ignites in an off-center shell (core helium
flash). No shell-shifting or other approximation is done on the RGB; the
full evolution is followed. Typically, the calculations need about 200
time-steps until core hydrogen exhaustion, another 300 until the onset of the
first dredge-up, 700 to the end of it and a further 8000 to the tip of
the RGB.

The spatial resolution of the models is such that on the main sequence
of order 600 and on the RGB twice as many grid-points are needed. We
verified that increasing the number of grid-points and time-steps does
not influence the relation between luminosity and age by more than a
per cent.

\subsection{Evolutionary tracks}

\begin{figure}[ht]
\centerline{\includegraphics[scale=0.80,bb= 30 0 420 280,draft=false]{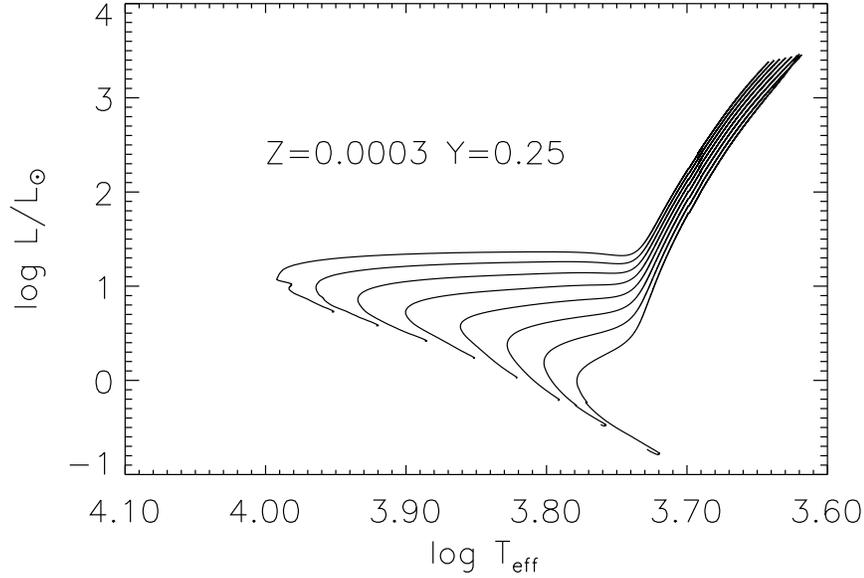}}
\caption[]{Evolution (without diffusion) in the HRD for all masses
($0.6,\,0.7,\ldots1.3\;M_\odot$) with composition $Y=0.25$, $Z=3\cdot10^{-4}$}
\protect\label{f:1}
\end{figure}

\begin{figure*}[ht]
\begin{center}	
\centerline{\includegraphics[scale=0.80,draft=false]{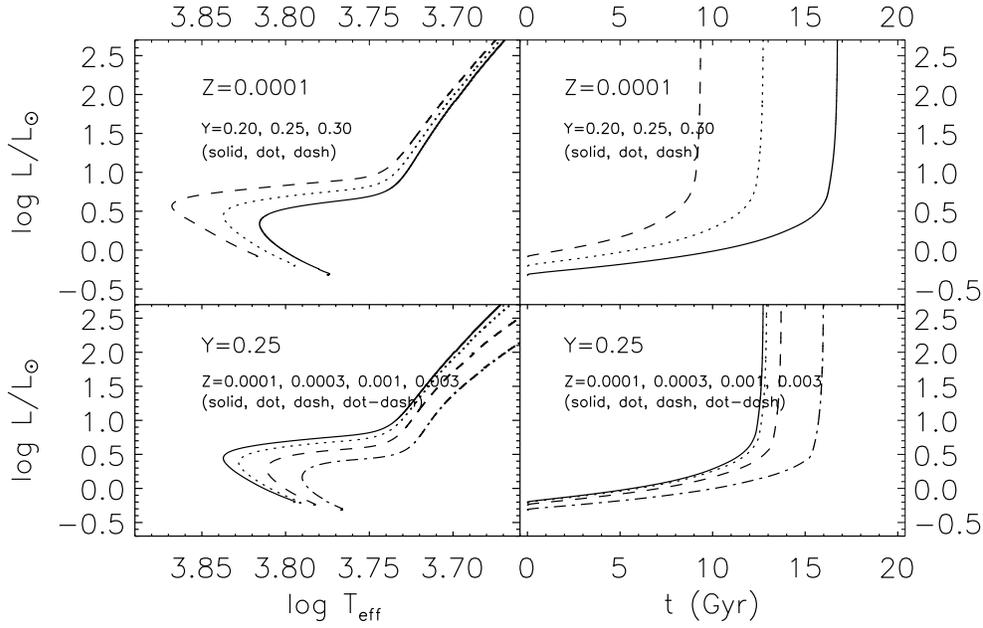}}
\caption[]{Influence of composition changes on the evolution of the
$0.8\,M_\odot$ model (no diffusion)} 
\protect\label{f:2}
\end{center}
\end{figure*}

We display in Fig.~\ref{f:1} the evolution without diffusion
(``C''-set) of all masses for the
case which is close to the centre of our 3x4 composition space, i.e.\ for
$(Y,\,Z) = (0.25,\,0.0003)$, and in Fig.~\ref{f:2} the changes of the
evolution due to variations of the composition for the case
of the $0.8\,M_\odot$ model. The left panels show the HRD-tracks
(top: varying helium content; bottom: varying metallicity) and the
right ones the evolutionary speed. The well-known effects, such
as a lower effective temperature for higher $Z$ or lower $Y$ or shorter
main-sequence (MS) lifetimes for higher $Y$ or lower $Z$, are recognizable.

\begin{figure*}[ht]
\begin{center}	
\centerline{\includegraphics[scale=0.70,draft=false]{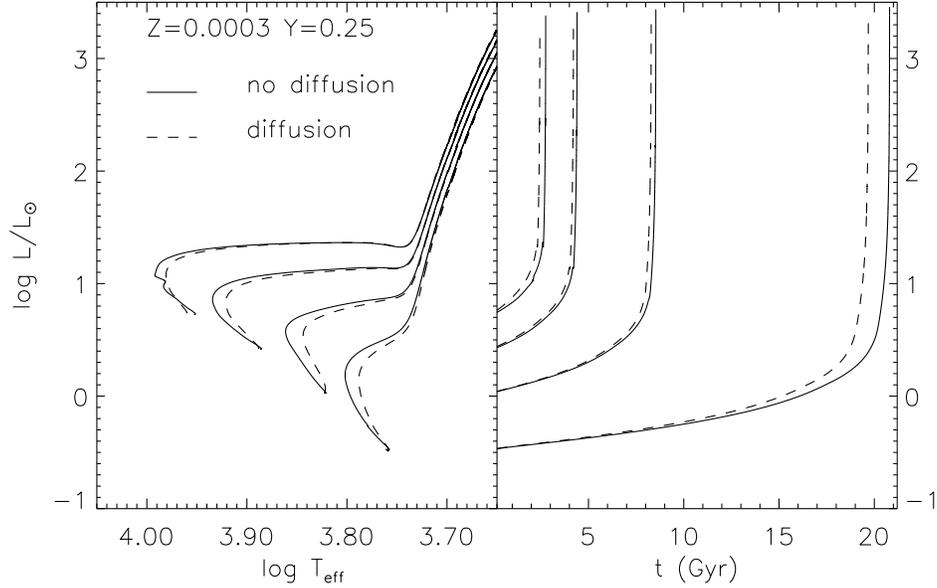}}
\caption[]{Influence of diffusion on the evolutionary tracks (left panel) and
lifetimes (right panel) for
selected masses ($0.7,\,0.9,\,1.1,\,1.3\,M_\odot$) and the same
composition as in Fig.~\ref{f:1}} 
\protect\label{f:3}
\end{center}
\end{figure*}

\begin{figure}[ht]
\mbox{~}\hspace{0.8cm}
\includegraphics[scale=0.45,angle=90,draft=false]{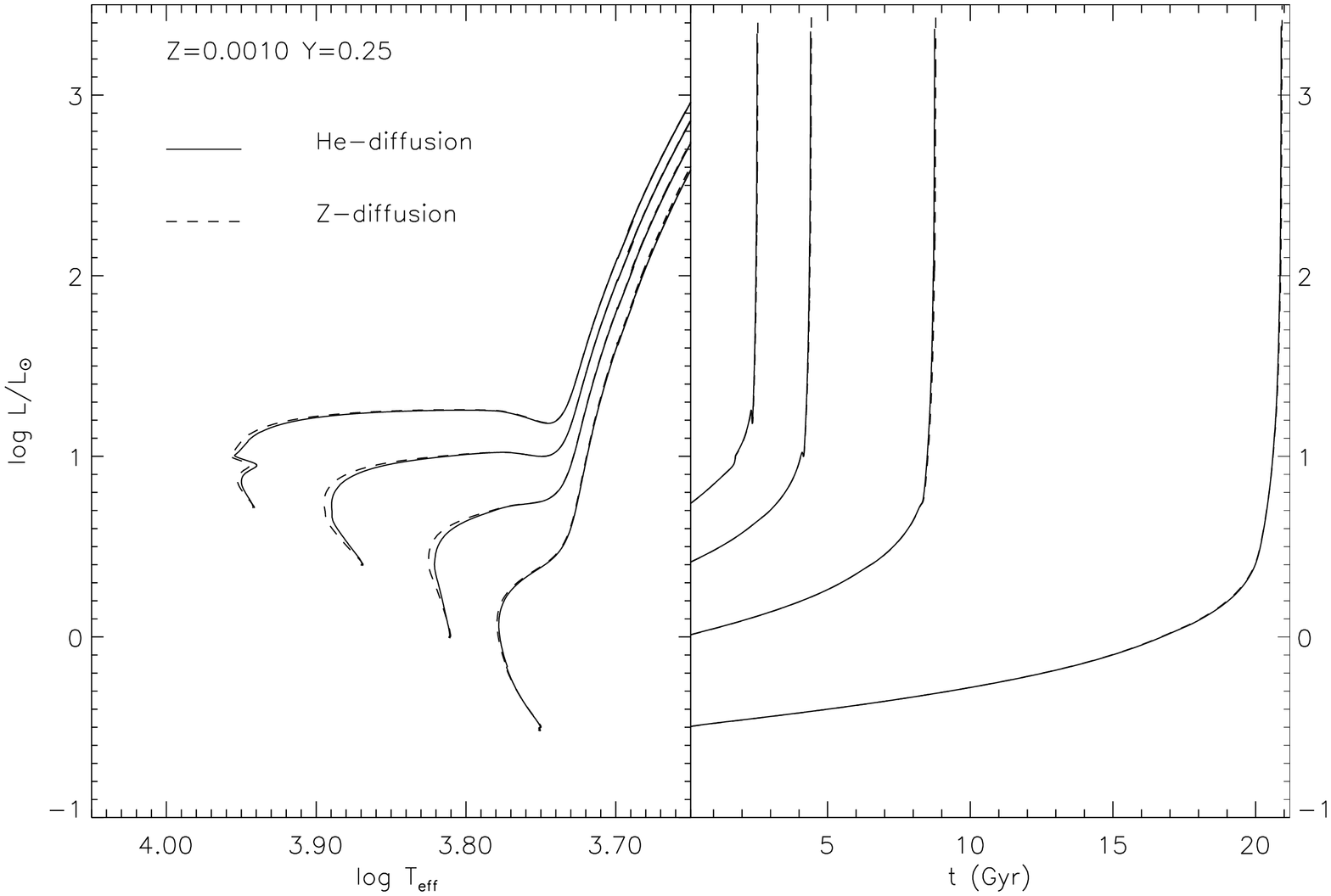}
\caption[]{Influence of metal diffusion on the evolutionary tracks (left panel) and
lifetimes (right panel) for composition  $(Y,\,Z)=(0.25,\,0.001)$ and
the same selected masses composition as in Fig.~\ref{f:3}. Comparison
is made with the case of hydrogen-helium diffusion} 
\protect\label{f:3a}
\end{figure}

The influence of diffusion both on the track in the HRD and on the
evolutionary speed is displayed in Fig.~\ref{f:3} for the same
reference composition. For sake of clarity the evolution of only a few
selected masses are shown. The effects -- for example, lower effective
temperature and brightness during the main sequence -- are as known
from other investigations (e.g.\ \cite{ccd:98}). MS-lifetimes get
shorter due to the diffusion of helium into the center, which is
effectively equivalent to a faster aging of the star. For given
MS-luminosity, TO-models with diffusion can be younger by up to 1~Gyr
compared to those calculated canonically. We recall that we include
only H/He-diffusion in the grid of models of this paper. To verify
that the additional metal diffusion has a negligible influence on the
evolutionary tracks and in particular on lifetimes, we show in
Fig.~\ref{f:3a} the comparison between models with H/He- and
H/He/Z-diffusion in the case of mixture $(Y,\,Z)=(0.25,\,0.001)$. We
chose a higher metallicity than in the previous example because the
depletion of the stellar envelopes in metals due to diffusion is
expected to have a higher effect for higher initial metallicity.  As
Fig.~\ref{f:3a} demonstrates, the age-luminosity relation is almost
identical and the track in the HRD only slightly shifted to the blue
because of the decrease in surface metallicty. After the turn-off, the
deepening convective envelope is mixing back quickly the diffused
elements such that the initial envelope composition is almost restored
(cf.\ \cite{sgw:00}). The tracks approach each other therefore during
the subgiant evolution. The surface metallicity drops to a minimum of
42\% of the initial one for the $M/M_\odot=0.8$ model, which is good
agreement with results by \citen{sgw:00}.

The evolutionary properties for {\em all} cases calculated are given
in tables in Appendix~A. 

\subsection{Age-luminosity relations}

\begin{figure}[ht]
\centerline{\includegraphics[scale=0.7,bb=30 20 420 290,draft=false]{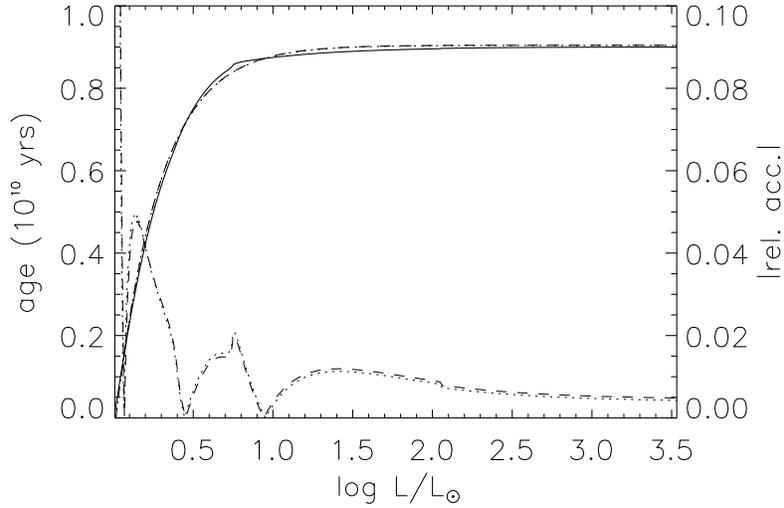}}
\caption[]{$t(\log L)$ (left axis) from the calculation of a (case C)
model with $M/M_\odot = 0.90$, $Y=0.25$, $Z=0.001$ (solid) and as
obtained from Eq.~(1) (dotted) or Eq.~(2) (dashed). The relative accuracy 
(absolute value) for both fitting functions is also shown
(corresponding thin lines; right axis)}
\protect\label{f:4}
\end{figure}

\begin{figure}[ht]
\centerline{\includegraphics[scale=0.70,bb=30 20 420 310,draft=false]{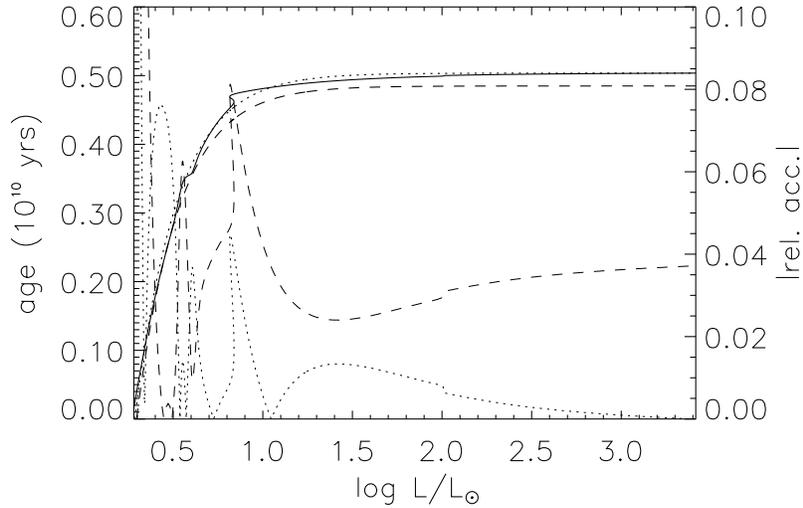}}
\caption[]{As Fig.~\ref{f:4}, but for a model with diffusion (case D)
and $M/M_\odot = 1.00$, $Y=0.30$, $Z=0.003$}
\protect\label{f:5}
\end{figure}

It would be desirable to have an analytical formula $t(L,M,Y,Z)$, which
returns the age of a star for any given set of observed
quantities. However, there is no simple analytical fit to the results of
the evolutionary calculations and high-order fitting formulae are not
practical. We have attempted to provide fits which are a compromise
between accuracy and simplicity and start with providing a fitting
formula to obtain $t(L)$ for each individual mass calculated. This
formula is
\begin{equation}
t{\rm [10^{10}~yr]} = a - \exp(-(\log(L/L_\odot)+b)\cdot c)
\end{equation}
\noindent and in general fits the evolutionary ages after the initial
100-200 Myr with an accuracy of a few percent. The coefficients $a$, $b$
and $c$ depend on mass and composition. Tables~\ref{t:fitc} and
\ref{t:fitd} contain the values of all of them for all cases. 
To illustrate the fit quality, both the fit and the relative fitting
accuracy are shown for two selected cases in Figs.~\ref{f:4} and \ref{f:5}
(solid vs.\ dotted lines).

\begin{table*}[ht]
\caption[]
{Parameters for the function $t= a -
\exp(-(\log(L/L_\odot)+b)\cdot c)$. $t$ is in $10^{10}$~years. For
each combination of  mass $M/M_\odot$, metallicity $Z$ and helium 
content $Y$ the fitting parameters $a$, $b$ and $c$ are given. Case
``C'' (no diffusion).}
\protect\label{t:fitc}
\begin{tabular}{cc|rrr|rrr|rrr}
$M/M_\odot$ & $Z$ & \multicolumn{3}{c|}{$Y=0.20$} &
\multicolumn{3}{c|}{$Y=0.25$} & \multicolumn{3}{c}{$Y=0.30$} \cr
& & \multicolumn{1}{c}{$a$} & \multicolumn{1}{c}{$b$} &
\multicolumn{1}{c|}{$c$} & \multicolumn{1}{c}{$a$} & \multicolumn{1}{c}{$b$} &
\multicolumn{1}{c|}{$c$} & \multicolumn{1}{c}{$a$} & \multicolumn{1}{c}{$b$} &
\multicolumn{1}{c}{$c$} \cr
\hline
 0.60   & 0.0001        & 4.68683  & 0.35586  & 3.05625 & 3.53605  & 0.34283  & 3.02336  & 2.60818  & 0.33038  & 3.02014 \\
        & 0.0003        & 4.77712  & 0.36200  & 3.06499 & 3.59575  & 0.34984  & 3.05363  & 2.65045  & 0.33699  & 3.05964 \\
        & 0.0010        & 5.10040  & 0.37619  & 3.07399 & 3.82210  & 0.36438  & 3.09124  & 2.81103  & 0.35093  & 3.11996 \\
        & 0.0030        & 6.06607  & 0.37026  & 2.94939 & 4.52971  & 0.38835  & 3.13366  & 3.24782  & 0.37750  & 3.16778 \\[0.3em]

 0.70   & 0.0001        & 2.75644  & 0.23948  & 3.07744 & 2.04816  & 0.22465  & 3.11651 & 1.50112  & 0.20670  & 3.16830 \\
        & 0.0003        & 2.80616  & 0.24708  & 3.12424 & 2.08362  & 0.23132  & 3.17035 & 1.52506  & 0.21224  & 3.22277 \\
        & 0.0010        & 2.98995  & 0.26352  & 3.19686 & 2.21790  & 0.24602  & 3.25090 & 1.61912  & 0.22531  & 3.30198 \\
        & 0.0030        & 3.49486  & 0.29429  & 3.25585 & 2.58532  & 0.27583  & 3.33550 & 1.88109  & 0.25322  & 3.39598 \\[0.3em]

 0.80   & 0.0001        & 1.68343  & 0.13525  & 3.22279 & 1.27581  & 0.11091  & 3.25539 & 0.93729  & 0.08773  & 3.30100 \\
        & 0.0003        & 1.74860  & 0.13805  & 3.27157 & 1.29541  & 0.11599  & 3.31912 & 0.94964  & 0.09107  & 3.37295 \\
        & 0.0010        & 1.86228  & 0.15350  & 3.37035 & 1.37588  & 0.12939  & 3.40681 & 1.00329  & 0.10161  & 3.46594 \\
        & 0.0030        & 2.18269  & 0.18574  & 3.48119 & 1.60682  & 0.15929  & 3.52297 & 1.16197  & 0.12978  & 3.56537 \\[0.3em]

 0.90   & 0.0001        & 1.13503  & 0.02607  & 3.34182 & 0.84599  & 0.00074  & 3.38876 & 0.62575 & -0.02647  & 3.43726 \\
        & 0.0003        & 1.15214  & 0.03091  & 3.41167 & 0.85656  & 0.00347  & 3.47214 & 0.63156 & -0.02646  & 3.53322 \\
        & 0.0010        & 1.22304  & 0.04434  & 3.51816 & 0.90404  & 0.01403  & 3.58107 & 0.66144 & -0.02009  & 3.66277 \\
        & 0.0030        & 1.43144  & 0.07579  & 3.65932 & 1.04797  & 0.04264  & 3.71174 & 0.75716 &  0.00348  & 3.81259 \\[0.3em]

 1.00   & 0.0001        & 0.78944  &-0.07630  & 3.47280 & 0.59243 & -0.10569  & 3.53159 & 0.44517 & -0.13915  & 3.61112 \\
        & 0.0003        & 0.79898  &-0.07378  & 3.56344 & 0.59726 & -0.10614  & 3.64016 & 0.44737 & -0.14257  & 3.73657 \\
        & 0.0010        & 0.84194  &-0.06398  & 3.70478 & 0.62523 & -0.10057  & 3.79452 & 0.46323 & -0.14090  & 3.91179 \\
        & 0.0030        & 0.97527  &-0.03456  & 3.87128 & 0.71489 & -0.07726  & 3.99567 & 0.51923 & -0.12586  & 4.15912 \\[0.3em]

 1.10   & 0.0001        & 0.57401  &-0.17656  & 3.62913 & 0.44051 & -0.21321  & 3.73425 & 0.34106 & -0.26244  & 3.91988 \\
        & 0.0003        & 0.57955  &-0.17673  & 3.74518 & 0.44161 & -0.21628  & 3.87160 & 0.34008 & -0.26682  & 4.06653 \\
        & 0.0010        & 0.60410  &-0.17304  & 3.95072 & 0.45619 & -0.21517  & 4.07949 & 0.34790 & -0.27014  & 4.34218 \\
        & 0.0030        & 0.69070  &-0.14894  & 4.19878 & 0.50901 & -0.19791  & 4.33119 & 0.37785 & -0.25579  & 4.66150 \\[0.3em]

 1.20   & 0.0001        & 0.44274  &-0.27904  & 3.85851 & 0.34658 & -0.32992  & 4.05838 & 0.26632 & -0.38467  & 4.26117 \\
        & 0.0003        & 0.44597  &-0.28260  & 4.01666 & 0.34535 & -0.33435  & 4.22578 & 0.26563 & -0.40273  & 4.59135 \\
        & 0.0010        & 0.45712  &-0.28112  & 4.25898 & 0.35288 & -0.33823  & 4.54620 & 0.27248 & -0.42019  & 5.10977 \\
        & 0.0030        & 0.50782  &-0.26315  & 4.55547 & 0.38208 & -0.32214  & 4.91328 & 0.28638 & -0.40417  & 5.49256 \\[0.3em]

 1.30   & 0.0001        & 0.35440  &-0.38785  & 4.17826 & 0.27521 & -0.44247  & 4.38679 & 0.21011 & -0.50132  & 4.62344 \\
        & 0.0003        & 0.35614  &-0.39527  & 4.39311 & 0.27396 & -0.45857  & 4.71954 & 0.21034 & -0.53311  & 5.13498 \\
        & 0.0010        & 0.36100  &-0.39740  & 4.74103 & 0.28144 & -0.47611  & 5.28857 & 0.21569 & -0.55507  & 5.76582 \\
        & 0.0030        & 0.39037  &-0.38075  & 5.16706 & 0.29669 & -0.46093  & 5.76533 & 0.22379 & -0.52555  & 5.86558 
\end{tabular}
\end{table*}

\begin{table*}
\caption[]
{As Table~\ref{t:fitc}, but for case ``D'' (with diffusion).}
\protect\label{t:fitd}
\begin{tabular}{cc|rrr|rrr|rrr}
$M/M_\odot$ & $Z$ & \multicolumn{3}{c|}{$Y=0.20$} &
\multicolumn{3}{c|}{$Y=0.25$} & \multicolumn{3}{c}{$Y=0.30$} \cr
& & \multicolumn{1}{c}{$a$} & \multicolumn{1}{c}{$b$} &
\multicolumn{1}{c|}{$c$} & \multicolumn{1}{c}{$a$} & \multicolumn{1}{c}{$b$} &
\multicolumn{1}{c|}{$c$} & \multicolumn{1}{c}{$a$} & \multicolumn{1}{c}{$b$} &
\multicolumn{1}{c}{$c$} \cr
\hline
 0.60 & 0.0001 & 4.32887 & 0.37567 & 3.02190 & 3.26232 & 0.36413 & 2.99265 & 2.41195 & 0.35193 & 2.99435 \\
      & 0.0003 & 4.40146 & 0.38369 & 3.04085 & 3.31250 & 0.37205 & 3.02861 & 2.44809 & 0.35903 & 3.03903 \\
      & 0.0010 & 4.68208 & 0.40064 & 3.06122 & 3.50928 & 0.38869 & 3.07645 & 2.58894 & 0.37435 & 3.10479 \\
      & 0.0030 & 5.49931 & 0.40621 & 2.96719 & 4.04219 & 0.41951 & 3.11132 & 2.96640 & 0.40530 & 3.16241 \\[0.3em]

 0.70 & 0.0001 & 2.60728 & 0.25458 & 3.07649 & 1.93855 & 0.23953 & 3.11674 & 1.42432 & 0.22087 & 3.16418 \\
      & 0.0003 & 2.65059 & 0.26277 & 3.12750 & 1.97077 & 0.24646 & 3.16889 & 1.44621 & 0.22648 & 3.22177 \\
      & 0.0010 & 2.82116 & 0.28025 & 3.20341 & 2.09415 & 0.26202 & 3.25551 & 1.53329 & 0.24023 & 3.30384 \\
      & 0.0030 & 3.27308 & 0.31450 & 3.26899 & 2.42215 & 0.29481 & 3.34407 & 1.76776 & 0.27068 & 3.41326 \\[0.3em]

 0.80 & 0.0001 & 1.64775 & 0.14155 & 3.22820 & 1.22635 & 0.12050 & 3.26636 & 0.90293 & 0.09644 & 3.31164 \\
      & 0.0003 & 1.67685 & 0.14833 & 3.28972 & 1.24557 & 0.12581 & 3.32984 & 0.91498 & 0.09989 & 3.38364 \\
      & 0.0010 & 1.78747 & 0.16442 & 3.38334 & 1.32252 & 0.13983 & 3.41672 & 0.96556 & 0.11174 & 3.47145 \\
      & 0.0030 & 2.11192 & 0.18609 & 3.33429 & 1.53360 & 0.17155 & 3.54847 & 1.11111 & 0.14061 & 3.58888 \\[0.3em]

 0.90 & 0.0001 & 1.10076 & 0.03244 & 3.36031 & 0.82064 & 0.00676 & 3.40366 & 0.60561 & -0.02245 & 3.46209 \\
      & 0.0003 & 1.11760 & 0.03745 & 3.43144 & 0.83141 & 0.00965 & 3.48631 & 0.61216 & -0.02207 & 3.55489 \\
      & 0.0010 & 1.18750 & 0.05159 & 3.53355 & 0.87804 & 0.02080 & 3.59323 & 0.64233 & -0.01466 & 3.68069 \\
      & 0.0030 & 1.38137 & 0.08447 & 3.66996 & 1.01169 & 0.05043 & 3.73441 & 0.73241 &  0.01050 & 3.82981 \\[0.3em]

 1.00 & 0.0001 & 0.76908 & -0.07288 & 3.49792 & 0.57533 & -0.10408 & 3.56717 & 0.42600 & -0.13838 & 3.63774 \\
      & 0.0003 & 0.77900 & -0.07044 & 3.59247 & 0.58121 & -0.10383 & 3.66964 & 0.42929 & -0.14256 & 3.77247 \\
      & 0.0010 & 0.82251 & -0.05927 & 3.72051 & 0.61003 & -0.09704 & 3.81815 & 0.44773 & -0.14085 & 3.95451 \\
      & 0.0030 & 0.94850 & -0.02913 & 3.89619 & 0.69552 & -0.07280 & 4.02281 & 0.50360 & -0.12202 & 4.19106 \\[0.3em]

 1.10 & 0.0001 & 0.55857 & -0.17708 & 3.67657 & 0.41930 & -0.21320 & 3.75835 & 0.31158 & -0.25374 & 3.86183 \\
      & 0.0003 & 0.56415 & -0.17722 & 3.79350 & 0.42210 & -0.21775 & 3.91291 & 0.31300 & -0.26297 & 4.04566 \\
      & 0.0010 & 0.59210 & -0.17067 & 3.97315 & 0.44043 & -0.21741 & 4.13788 & 0.32390 & -0.26772 & 4.30476 \\
      & 0.0030 & 0.67508 & -0.14611 & 4.22927 & 0.49579 & -0.19793 & 4.40999 & 0.36014 & -0.26203 & 4.79190 \\[0.3em]

 1.20 & 0.0001 & 0.41916 & -0.28013 & 3.88606 & 0.31552 & -0.32160 & 3.99623 & 0.23514 & -0.36652 & 4.11702 \\
      & 0.0003 & 0.42181 & -0.28496 & 4.05931 & 0.31673 & -0.33159 & 4.20871 & 0.23532 & -0.38079 & 4.35825 \\
      & 0.0010 & 0.43987 & -0.28553 & 4.33719 & 0.32785 & -0.33961 & 4.54623 & 0.24104 & -0.39874 & 4.84120 \\
      & 0.0030 & 0.49553 & -0.26471 & 4.64456 & 0.36493 & -0.33050 & 5.06159 & 0.26657 & -0.40618 & 5.60667 \\[0.3em]

 1.30 & 0.0001 & 0.32287 & -0.38193 & 4.13115 & 0.24368 & -0.42705 & 4.25755 & 0.18201 & -0.47279 & 4.36969 \\
      & 0.0003 & 0.32393 & -0.39226 & 4.37277 & 0.24378 & -0.44282 & 4.53974 & 0.17950 & -0.49064 & 4.69556 \\
      & 0.0010 & 0.33549 & -0.39965 & 4.75053 & 0.24998 & -0.45894 & 5.04221 & 0.18393 & -0.50652 & 5.21632 \\
      & 0.0030 & 0.37394 & -0.39030 & 5.34027 & 0.27690 & -0.46505 & 5.89274 & 0.20304 & -0.51820 & 6.03615 
\end{tabular}
\end{table*}

As a next step, we tried to model the dependence of the fitting
coefficients in Eq.~(1) on mass. Globally, $a(M)$ and $c(M)$ appear to
be reminiscent of a parabolic function, while $b(M)$ is close to a
linear one. However, individual coefficient values lie off the main
trend, such that higher order fits are required. We used the cubic
polynomial
\begin{equation}
a = a_0 + a_1\cdot (M/M_\odot) + a_2\cdot (M/M_\odot)^2 + a_3 \cdot
(M/M_\odot)^3 
\end{equation}
\noindent (and equivalent expressions for $b$ and $c$). The individual
coefficients still depend on $Y$ and $Z$. We found that with this
approximation the coefficients could be modelled again with an accuracy
of a few percent. The coefficients $a_0\ldots c_3$ are listed for both
sets of calculations in Tables~\ref{t:fitcm} and \ref{t:fitdm}. The fits
obtained by using Eq.~(2) with the coefficients taken from
Tabs.~\ref{t:fitcm} and \ref{t:fitdm} and inserting the proper mass value
to obtain the coefficients of Eq.~(1) are shown in Figs.~\ref{f:4}
and \ref{f:5} (dashed lines) as well. While in the former case the fit
quality is not degraded, it is worse in the latter one, but the errors
remain within 5\% for most of the evolution (see the thin dashed
line), except for the very end of
the main-sequence, where the luminosity of the more massive stars shows
the complicated ``kink''-behaviour in the HRD, which cannot be modelled
by this fitting function.

It is not useful to continue with finding fitting functions in
composition, because we have only a 3x4 composition space and such
functions would require at least 3 parameters for each
dimension. Further fitting would therefore only increase the total
number of coefficients. Rather, we recommend to interpolate between the
ages obtained through Eqs.~(2) and (1) to the composition of any
observed star.

\newpage

\begin{table}[ht]
\caption[]
{Fitting the mass-dependence of the coefficients of Table~\ref{t:fitc}. The
three lines of each composition correspond to the three 
parameters a, b, and c; for example, the first lines contain
coefficients $a_0$, $a_1$, $a_2$, and $a_3$ of Eq.~(2).}
\protect\label{t:fitcm}
\begin{tabular}{rr|rrrr}
$Y$ & $Z$ & \multicolumn{4}{c}{a, b, c} \\
\hline
0.20 & 0.0001 &     29.2077 &    -70.7155 &     58.7312 &    -16.4698 \\
     &        &      1.4060 &     -2.4031 &      1.3708 &     -0.4489 \\
     &        &      0.1911 &      9.0341 &     -9.6597 &      3.8994 \\[0.3em]
0.20 & 0.0003 &     29.7043 &    -71.7262 &     59.3967 &    -16.6087 \\
     &        &      1.3975 &     -2.3636 &      1.3456 &     -0.4525 \\
     &        &     -0.2651 &     10.6190 &    -11.4234 &      4.6238 \\[0.3em]
0.20 & 0.0010 &     31.4324 &    -75.6675 &     62.4585 &    -17.4163 \\
     &        &      1.3120 &     -2.0165 &      0.9768 &     -0.3362 \\
     &        &     -1.5348 &     14.8288 &    -15.9140 &      6.3227 \\[0.3em]
0.20 & 0.0030 &     36.8363 &    -88.4998 &     72.8976 &    -20.2998 \\
     &        &      0.8201 &     -0.4514 &     -0.5362 &      0.1330 \\
     &        &     -4.3950 &     23.6936 &    -24.8876 &      9.4754 \\[0.3em]
0.25 & 0.0001 &     22.5447 &    -55.0764 &     46.1447 &    -13.0444 \\
     &        &      1.1258 &     -1.5493 &      0.4975 &     -0.1800 \\
     &        &      2.7059 &      1.1295 &     -1.7242 &      1.4248 \\[0.3em]
0.25 & 0.0003 &     22.8452 &    -55.7195 &     46.5951 &    -13.1477 \\
     &        &      1.4040 &     -2.4222 &      1.4181 &     -0.5054 \\
     &        &     -1.0202 &     13.5479 &    -15.1164 &      6.2240 \\[0.3em]
0.25 & 0.0010 &     24.0297 &    -58.3325 &     48.5215 &    -13.6191 \\
     &        &      1.6165 &     -3.0661 &      2.1179 &     -0.7673 \\
     &        &     -4.8060 &     26.6803 &    -29.9153 &     11.8184 \\[0.3em]
0.25 & 0.0030 &     27.9680 &    -67.6242 &     55.9981 &    -15.6582 \\
     &        &      1.4430 &     -2.3857 &      1.3930 &     -0.5264 \\
     &        &     -4.6154 &     26.4036 &    -29.9896 &     12.1702 \\[0.3em]
0.30 & 0.0001 &     16.8533 &    -41.4895 &     35.0619 &     -9.9976 \\
     &        &      0.8082 &     -0.5057 &     -0.6429 &      0.1976 \\
     &        &      4.7946 &     -5.7016 &      5.4242 &     -0.8754 \\[0.3em]
0.30 & 0.0003 &     17.0790 &    -41.9770 &     35.3982 &    -10.0704 \\
     &        &      0.9524 &     -1.0040 &     -0.0459 &     -0.0471 \\
     &        &      3.5676 &     -0.6182 &     -1.2110 &      2.0137 \\[0.3em]
0.30 & 0.0010 &     18.1267 &    -44.4785 &     37.4060 &    -10.6092 \\
     &        &      0.7120 &     -0.1966 &     -0.8545 &      0.1964 \\
     &        &      7.8855 &    -13.4807 &     10.8945 &     -1.3620 \\[0.3em]
0.30 & 0.0030 &     20.4666 &    -49.7150 &     41.3637 &    -11.6164 \\
     &        &      0.3008 &      1.3422 &     -2.5921 &      0.8232 \\
     &        &     20.2151 &    -54.5229 &     55.3097 &    -16.8058 \\
\end{tabular}
\end{table}

\begin{table}[ht]
\caption[]{As Tab.~\ref{t:fitcm}, but the case D coefficients of
Tab.~\ref{t:fitd}} 
\protect\label{t:fitdm}
\begin{tabular}{rr|rrrr}
$Y$ & $Z$ & \multicolumn{4}{c}{a, b, c} \\
\hline
0.20 & 0.0001 &     26.4015 &    -63.2538 &     52.1585 &    -14.5656 \\
     &        &      1.2280 &     -1.6948 &      0.5350 &     -0.1415 \\
     &        &      2.1368 &      2.1647 &     -1.8484 &      1.0488 \\[0.3em]
0.20 & 0.0003 &     26.7886 &    -64.1169 &     52.8122 &    -14.7334 \\
     &        &      1.2478 &     -1.7368 &      0.5949 &     -0.1764 \\
     &        &      1.6609 &      3.9792 &     -4.0180 &      1.9709 \\[0.3em]
0.20 & 0.0010 &     28.3804 &    -67.7799 &     55.6983 &    -15.5053 \\
     &        &      1.2038 &     -1.5540 &      0.4377 &     -0.1471 \\
     &        &      1.7792 &      4.1186 &     -4.7200 &      2.5472 \\[0.3em]
0.20 & 0.0030 &     32.9185 &    -78.2930 &     64.0441 &    -17.7525 \\
     &        &      1.1598 &     -1.4451 &      0.4589 &     -0.2035 \\
     &        &     -4.9779 &     25.6408 &    -27.2815 &     10.5084 \\[0.3em]
0.25 & 0.0001 &     19.9386 &    -47.9299 &     39.6462 &    -11.1021 \\
     &        &      1.1975 &     -1.5798 &      0.3577 &     -0.0798 \\
     &        &      2.2408 &      1.7358 &     -1.2970 &      0.8889 \\[0.3em]
0.25 & 0.0003 &     20.2355 &    -48.6066 &     40.1675 &    -11.2377 \\
     &        &      1.1670 &     -1.4564 &      0.2420 &     -0.0571 \\
     &        &      2.5421 &      1.0307 &     -0.8450 &      0.9498 \\[0.3em]
0.25 & 0.0010 &     21.4126 &    -51.3506 &     42.3545 &    -11.8293 \\
     &        &      1.0981 &     -1.1875 &     -0.0110 &      0.0024 \\
     &        &      2.0184 &      3.4628 &     -4.2511 &      2.5979 \\[0.3em]
0.25 & 0.0030 &     24.4112 &    -58.2151 &     47.7292 &    -13.2546 \\
     &        &      1.2151 &     -1.5061 &      0.3984 &     -0.1801 \\
     &        &     -1.2354 &     15.6116 &    -18.9391 &      8.5766 \\[0.3em]
0.30 & 0.0001 &     14.7503 &    -35.5174 &     29.4317 &     -8.2556 \\
     &        &      1.0895 &     -1.2037 &     -0.1142 &      0.0889 \\
     &        &      3.4016 &     -2.1258 &      2.9171 &     -0.5447 \\[0.3em]
0.30 & 0.0003 &     15.0295 &    -36.2278 &     30.0487 &     -8.4381 \\
     &        &      1.0026 &     -0.8747 &     -0.4748 &      0.2030 \\
     &        &      3.2613 &     -1.4881 &      2.0549 &     -0.0469 \\[0.3em]
0.30 & 0.0010 &     15.8313 &    -38.0600 &     31.4652 &     -8.8065 \\
     &        &      0.5327 &      0.7118 &     -2.1346 &      0.7475 \\
     &        &      9.0922 &    -19.5958 &     20.0909 &     -5.6196 \\[0.3em]
0.30 & 0.0030 &     18.0913 &    -43.3256 &     35.6464 &     -9.9272 \\
     &        &      0.0508 &      2.3708 &     -3.8357 &      1.2883 \\
     &        &     20.6253 &    -56.0962 &     57.0595 &    -17.3318 \\
\end{tabular}
\end{table}
\clearpage

\section{Discussion}

\subsection{Comparison with other theoretical results}

\citen{pacz:97} has argued that the achievable accuracy in the
observations of detached eclipsing binaries in globular clusters 
translates into an accuracy of the determined stellar age of order
2\%. Very justified, Paczy\'nski states that ``the uncertainties in the
stellar models are certainly larger than that''. To give an impression
of how large these uncertainties might be, we compare our results to some
other contemporary calculations of comparable models.

In a first step we compare ZAMS positions. \citen{tpe:96} have given
analytic functions for ZAMS-positions as function of mass and
metallicity based on their own calculations. Since they used a fixed
$Y$-$Z$-relation we cannot straightforwardly compare their results with
our low-metallicity models. The composition closest to their relation is that
with $Y=0.25$ and $Z=0.0010$, which is to be compared with $Y=0.2420$ (and
the same $Z$). We find that over the mass range of our calculations our
ZAMS models are 0.04~dex brighter in $L$ (with very
small variation) and slightly hotter (0.04~dex in $T_{\rm eff}$
for the lowest masses to 0.01~dex for the highest ones). Both effects
are consistent with the higher helium content of our models and the
fact that we are using a more up-to-date EOS. For a
solar-like mixture ($Y=0.28$, $Z=0.02$), for which we made additional
calculations, the differences are below the
0.02~dex level, reflecting the EOS-change only. We have also
calculated a set of ZAMS models for all 
three metallicities and the same helium content as in \citen{tpe:96},
but using our old EOS. In this case differences are below 0.01~dex
both in luminosities and effective temperatures with no systematic
effect recognizable. 

\begin{table}[ht]
\caption[]{Comparison between our models and those of \citen{bca:97} with
$Z=0.0002$ (upper group) and $0.001$ (lower group). The first line in
each case (0.7 and 0.8 $M/M_\odot$) 
gives the \citen{bca:97} data at 10~Gyr, the second and third line our
results for the two bracketing metallicities resp.\ the second line
our corresponding model. The comparison is made
at same age (colums 2 and 3) or at same luminosity (columns 4 and 5;
age in Gyr). All models have $Y=0.25$.}
\protect\label{t:isab}	
\begin{tabular}{l|rr|rr}
$M$/$Z$         & $\lg T_{\rm eff}$ & $\log L/L_\odot$ & $\lg T_{\rm
eff}$ & age \\
\hline
$M/M_\odot=0.8$ & 3.825 & 0.334 \\
$Z=0.0001$      & 3.832 & 0.285 & 3.835 & 10.4 \\
$Z=0.0003$      & 3.825 & 0.255 & 3.828 & 10.8 \\
\hline
$M/M_\odot=0.7$ & 3.772 & -0.265 \\ 
$Z=0.0001$      & 3.784 & -0.233 & 3.781 & 8.8 \\
$Z=0.0003$      & 3.780 & -0.250 & 3.779 & 9.4 \\
\hline
$M/M_\odot=0.8$ & 3.800 &  0.199 \\
$Z=0.001$       & 3.809 &  0.165 & 3.810 & 10.4 \\
\hline
$M/M_\odot=0.7$ & 3.755 & -0.326 \\
$Z=0.001$       & 3.771 & -0.298 & 3.768 & 8.8  \\
\end{tabular}
\end{table}

Next, we compared with results by \citen{bca:97}, where the most
important difference to our calculations is the use of the 
Saumon-Chabrier EOS (\cite{sch:95}). \citen{bca:97} list data for a
set of models with $Y=0.25$ and metallicities of $Z=0.001$ and
$0.0002$, the latter one being intermediate between the lower two of our
values. The age of these models is 10~Gyr. Table~\ref{t:isab} shows
how our models compare either at the same age or for the same
luminosity. Comparing at the same age, our models are less luminous by
about 0.06~dex for $M/M_\odot=0.8$ but brighter by about 0.02~dex for
$M/M_\odot=0.7$. This translates into age differences (if comparison at identical
luminosity is made) of about $+0.5$ resp.\ $-1.0$~Gyr for both
metallicities. \citen{bca:97} used solar metal ratios, but at these low absolute
metallicities there is almost no 
influence of the internal metal composition (solar or
$\alpha$-enhanced) as verified by \citen{bca:97} themselves (see also
\cite{sw:98}).

\begin{table*}
\caption[]{Comparison between turn-off data for our models and those
of \citeau{ccd:98} (\citeyear{ccd:98}; their Tab.~2) and \citen{gbbc:99}.
All models have $Y=0.23$. } 
\protect\label{t:franec}	
\begin{tabular}{ll|rrr|rrr|rrr}
$Z$ & $M/M_\odot$ & age &  $\log L/L_\odot$ & $\lg T_{\rm eff}$ & 
                  age &  $\log L/L_\odot$ & $\lg T_{\rm eff}$  & 
                  age &  $\log L/L_\odot$ & $\lg T_{\rm eff}$\\
\hline
& & \multicolumn{3}{c|}{Cassisi et al.~(1998)} 
  & \multicolumn{3}{c|}{Girardi et al.~(1999)} 
  & \multicolumn{3}{c}{this paper} \\
\hline
0.0001 &0.80 &11.6&0.410&3.826&&&&12.2&0.415&3.829\\
0.0002 &0.80 &11.2&0.378&3.824&&&&12.1&0.380&3.824\\
0.0010 &0.70 &20.0&0.060&3.777&21.6&0.074&3.784&21.8&0.074&3.783\\
0.0010 &0.80 &11.7&0.231&3.799&12.4&0.240&3.807&12.4&0.249&3.805\\
0.0010 &0.90 & 7.4&0.393&3.822&7.6&0.406&3.829&7.7&0.394&3.828\\
0.0010 &1.00 & 5.0&0.577&3.852&5.0&0.567&3.855&5.5&0.586&3.857\\
\end{tabular}
\end{table*}

As a further test, we compared with results obtained with the FRANEC code, in
particular those by \citen{ccd:98}, who provide results for several combinations
of input physics data. Their case-8 models are very similar to ours with the
major exception being the treatment of the EOS outside the OPAL-range. We
compare turn-off (TO) data for several cases in Tab.~\ref{t:franec}. The helium
content is 0.23 for all models; we have made additional calculations
with the same metallicity for this
purpose. They were done without an explicit network (equilibrium abundances for
the participating nuclei assumed). An explicit network increases the TO-age of the
$0.8\,M/M_\odot$ model ($Z=0.001$) by 0.5~Gyr; the inclusion of both network and
pre-main sequence phase results in an increase of only 0.3~Gyr. We add that the
calculations of \citen{sw:98}, done with a variant of the FRANEC code, produce
practically the same results as those by \citen{ccd:98}; the small differences can be
traced back mainly to the slightly higher helium content of 0.233 (at
$Z=0.001$). Overall, our models take longer to finish the MS-phase, with the
differences getting smaller for higher metallicity and mass. With one exception,
TO-ages are larger by less than 1~Gyr, or 5-10\%. The comparison has been
extended for RGB-tip data recently by \citen{cdgm:00}, finding similar
agreement. In the same table, for the highest metallicity, data from the latest
Padua-tracks (\cite{gbbc:99}) are listed as well. In this case, the agreement
with our own results is even better.

We close this part with a few remarks on comparing
isochrones with those by \citeau{dcm:97} (\citeyear{dcm:97}; Tab.~2), who
provide turn-off data for a large 
number of metallicities. At $Z=0.0002$ and $Y=0.23$, for which we
again have made separate calculations, our 12~Gyr isochrone's TO is at
$\log L/L_\odot = 0.469$ and $\log T_{\rm eff} = 3.829$, which is
0.021~dex brighter and 0.03~dex hotter than the corresponding one by
\citen{dcm:97} (for mixing-length theory convection). The turn-off mass of
$0.820 M_\odot$ is larger by $0.012 M_\odot$. For $Z=0.001$, the differences
are very similar ($\delta\log L/L_\odot = 0.031$ and a TO-mass higher
by $0.018 M_\odot$). In fact, our TO-values are very close to those of
the 11~Gyr isochrone of \citen{dcm:97}. Part of the difference can be
ascribed to different helium abundances, which is $Y=0.235$ for their
models in this case. A similar comparison with the \citen{sw:98}
isochrones (for $Y=0.233$ and $Z=0.001$) gave an almost identical
result: while at 9~Gyr our isochrone is very close to their corresponding one,
the TO-brightness of our 13~Gyr isochrone is almost 
coincident with the 12~Gyr one of \citen{sw:98}. This result is
naturally to be expected from the comparison of Tab.~\ref{t:franec}.
To conclude, it appears that the different low-mass star calculations
agree with each other rather well, but the remaining differences,
which are partly due to physical assumptions and partly due to
technical details translate into age differences of up to 10\% for any
given composition, mass and luminosity. This can be viewed as the
inherent uncertainty the evolutionary calculations carry with them.

\subsection{A test application}

The intended application of our tracks are detached eclipsing binary
systems in globular clusters, which have been detected mainly by the
OGLE team in several clusters (\cite{kks:96};
\cite{kks:97}). Presently, for none of them follow-up  
spectroscopy needed to determine absolute parameters, has been
concluded, although for one system in $\omega$~Cen preliminary
data have been obtained (Kaluzny, private communication). Neither is
there any other suitable system from another 
source available. All well-known systems, e.g.\ CM~Dra and YY~Gem
(\cite{cb:95}), $\mu$~Cas (\cite{lpc:99}), or Gl570BC (\cite{fbd:99}),
are too metal-rich (${\rm [Fe/H]} > -0.76$). We therefore turned to
appropriate single stars to apply our relations and
tracks. \citen{fuhr:98} provides a list of nearby disk and halo stars
for which absolute parameters have been derived from a careful
spectroscopic analysis in conjunction with Hipparcos parallaxes. From
this list we have selected the five most metal-poor stars, of which
two, however, are slightly beyond the upper boundary of our metallicity
range (Tab.~\ref{t:fuhrdw}). All stars are enriched in Mg ($0.28 \le 
{\rm [Mg/Fe]} \le 0.46$) and
therefore are assumed to be $\alpha$-enriched in agreement with our
model compositions. Errors in $\rm [Fe/H]$ and $M_{\rm bol}$ are given
in Tab.~\ref{t:fuhrdw} as well, and are usually very small. The
largest uncertainty comes from the mass, which \citen{fuhr:98}
estimates to be of order 5\%, or generally, less than 10\%. The
uncertainty in $T_{\rm eff}$ is $\pm 80$~K in all cases. 
\citen{fuhr:98} classifies 4 of the selected stars as halo stars, and
the fifth one (HD201891) as belonging to the thick disk.

If atomic diffusion is in operation, the
presently observed and spectroscopically determined metallicity 
depends on both the initial one and on age. 
While in globular clusters the initial metallicity of main-sequence
and turn-off stars can be estimated
quite accurately from that of cluster giants (\cite{sgw:00}), this is
not possible for field stars. The degeneracy mentioned therefore
does not allow to determine the age independently of some assumptions
about the initial metallicity. We therefore applied only our
$t(L)$-fitting formulae Eq.~(1) and (2) 
without diffusion (case C) to these objects. 

Table~\ref{t:fuhrdw} contains age estimates in three steps: Column 6 gives the
age derived from the models with $Y=0.25$ and a metallicity closest to the
determined one (column 2), i.e.\ without any interpolation in $\rm
[Fe/H]$ (cf.\ Tab.~\ref{t:fehv}). In column 8, the age
obtained from interpolation to the observed $\rm [Fe/H]$ (but the same
helium content) is listed, and in column 10 that resulting from
interpolation to $Y=0.235$ (the ``generic'' Pop~II helium
abundance).

\begin{table*}[ht]
\caption[]{Application of Eq.~{2} (for the ``C''-case without
diffusion) to dwarfs from the sample of
\citen{fuhr:98} (columns 1--5). Errors according to the original paper
are given in the second line for each object.
Ages are derived in three steps (groups (1)--(3)): The first one using
a mixture with $Y=0.25$ and the metallicity closest to that of the
object (cf.\ Tab.~\ref{t:fehv}). The second step is to interpolate to
the observed metallicity, but still assuming $Y=0.25$. Finally, full
interpolation to the object's metallicity and $Y=0.235$ is done. For
this case, the age uncertainties due to the errors in metallicity,
bolometric magnitude and mass ($\triangle t_1$--$\triangle t_3$) are
given. All ages (in $10^9$~yrs) were obtained from application of
Eq.~(2), while theoretical effective temperatures, given next to ages,
were obtained from the evolutionary tracks. 
Explanation of remarks: (1) age taken directly from evolutionary tracks:
step (1): 8.79; step (3): 10.43 Gyr; (2)
$T_{\rm eff}$ taken from $0.6\,M_\odot$-tracks only, since $0.7\,M_\odot$
is always brighter than observed $M_{\rm bol}$; (3) no $T_{\rm eff}$
derived, {\rm [Fe/H]} being too high.}
\label{t:fuhrdw}
{\tiny
\begin{tabular}{l|rrrr|rr|rr|rrrrr|l}
{\bf object} & \multicolumn{4}{c|}{\bf stellar parameters} &
\multicolumn{2}{c|}{\bf (1)} & \multicolumn{2}{c|}{\bf (2)} &
\multicolumn{5}{c|}{\bf (3)} &  \cr
& $\rm [Fe/H]$ & $M_{\rm bol}$ & $M_\odot$ & $T_{\rm eff}$ &
age & $T_{\rm eff}$ & age &
$T_{\rm eff}$ & age & $T_{\rm eff}$ & $\triangle t_1$&$\triangle t_2$
& $\triangle t_3$  & note \\ 
\hline 
HD19445 & -1.95 & 4.91 & 0.74 & 6016 & 10.8 & 6289 & 11.3 &
6243 & 13.4 & 6122 & $\pm 0.2$ & $\pm 1.0$ & 9.9--18.0 & \\
        & 0.07 & 0.11 & 0.037 & 80 & & & & & & & & & 12.5--14.3\\
HD45282 & -1.52 & 1.98 & 0.90 & 5282 &   8.9 & 5321 &  9.2 &
5218 & 10.2 & 5267 & $\pm 0.2$ & $\pm 0.1$ & 8.3--12.6 & 1\\
        & 0.06 & 0.31 & 0.045 & 80 & & & & & & & & & 9.8--10.6 &\\
HD103095& -1.35 & 6.33 & 0.64 & 5110 &   8.9 & 5128 &  5.8 &
5195 &  9.7 & 5184 & $\pm 1.5$ & $\pm 1.6$ & $<$ 18.1 & 2\\
        & 0.10 & 0.05 & 0.032 & 80 & & & & & & & & & &\\
HD194598& -1.12\ & 4.45 & 0.84 & 6058 &   9.3 & 6232 &  9.4 &
6230 & 11.1 & 6151 & $\pm 0.4$ & $\pm 1.0$ & 7.7--15.0\\
        & 0.07 & 0.16 & 0.042 & 80 & & & & & & & & & 10.3--11.9 & \\
HD201891& -1.05 & 4.46 & 0.81 & 5943 &  11.9 & --- & 12.4 & ---
& 14.4 & --- & $+1.4$ & $\pm 0.5$ & 10.5--18.7 &3 \\
        & 0.08 & 0.09 & 0.041 & 80 & & & & & & & & & &\\
\end{tabular} }
\end{table*}

The derived ages appear to be rather consistent except for the 5.8~Gyr
for HD103095 (step 2), which is the most unevolved and least massive
object (see Fig.~2 of \cite{fuhr:98}). The final ages ($Y=0.235$) range
from 9.7--14.4 and are therefore in rough agreement with cluster ages
(\cite{sw:98}) computed with similar models. In the case of HD45282,
we can derive the age also directly from the evolutionary tracks for this mass
($0.9\,M_\odot$), i.e.\ without employing our fit formulae. We then
obtain for step 1 8.79~Gyr (compared to 8.9 from Eq.~(2); column 6)
and 10.43~Gyr compared to 10.2~Gyr (column 10) for 
the final mixture interpolation. This emphasizes the negligible error
due to Eqs.~(1) and (2) for these typical ages. In general, we find
that ages obtained by linear interpolation between the tracks agree
with the fitting formulae results to 5\% or better.

Columns 12--14 of Tab.~\ref{t:fuhrdw} list the age uncertainties
resulting from the errors in the observational quantities, which are
given in the second lines of columns 2--5. Obviously, the mass
uncertainty of 5\% (assumed for column 4) is by far too large to allow
accurate age determinations. The resulting age range ($\triangle t_3$)
is of order 8~Gyr, especially for the lower masses.  This emphasizes
the need for evolved objects close to or after the turn-off. HD45282
is such an object, beginning already its RGB ascent. For three objects
we give $\triangle t_3$ under the assumption that the mass is accurate
to 1\% (second line of $\triangle t_3$), which would lead to
acceptable uncertainties. This is also the achievable accuracy in
detached eclipsing binary systems (\cite{pacz:97}).  The age
uncertainty due to metallicity ($\triangle t_1$) is almost negligible
and that due to brightness -- i.e.\ distance -- errors ($\triangle
t_2$) of order 1~Gyr or smaller.  An exception is HD103095, which, due
to its low mass and unevolved state is of course most sensitive. In
this case, the upper mass limit of $0.68\, M_\odot$ is actually
inconsistent with the lowest (zero-age) brightness of our stellar
models.  A lower limit for $\triangle t_3$ is therefore missing.
HD201891 is outside the metallicity range of our models; its age of
14.4~Gyr, which is the highest of all objects, might be the result of
applying Eq.~(2) outside its definition range. The upper limit of the
metallicity range (${\rm [Fe/H]}=-0.97$) was not explored.  For
HD194598 the small extrapolation was allowed.

As a further consistency test we derived effective temperatures by
interpolating between the tracks. These $T_{\rm eff}$ are always given
in the column following that with the age.  The agreement with the
observed temperatures is, at least in the final case ($Y=0.235$) of
order of the $T_{\rm eff}$-error, with a tendency, however, that our
temperatures are higher. This could be an indication that diffusion,
which has been ignored here, is indeed active (\cite{sgw:00}).
$T_{\rm eff}$ for HD103095 was derived from the $0.6\, M_\odot$ tracks
only for the reasons given in the previous paragraph.

We finally comment on the use of models including diffusion. {\em
Assuming} that the typical metal depletion for a low-mass star of
cosmological age is of order 0.3~dex (\cite{sgw:00}), we have applied
the $t(L)$-relation of our D-models to HD19445 and HD194598. Then the
final (step 3) ages turn out to be 13.6 and 10.7~Gyr, which is
slightly older than in the C-case. Effective temperatures are reduced
to 5931 resp.\ 5194~K. Both values are again within the observational
uncertainties.

\subsection{Conclusions}

The comparison with other calculations and the application to (single)
stars with determined absolute stellar parameters revealed that the
largest errors in age determinations based on our stellar evolution
tracks are (1) mass, which must be known to 1\% accuracy and (2)
systematic uncertainties/differences in and between theretical models.

A physical source of uncertainty concerns the effectiveness of diffusion. In our
D--calculations, full diffusion of hydrogen and helium with coefficients calculated
following \citen{tbl:94} was included. This leads in many cases, due to the
extremly thin convective envelopes of metal-poor main-sequence stars, to an
almost complete depletion of the models in helium, which accumulates below the
convective layers. As soon as the star gets cooler, the convective envelope
deepens and the helium is mixed back to the surface, as is reflected in
the vanishing differences in the HRD in Fig.~\ref{f:3}\footnote{Very low helium
abundances towards the end of the main sequence are found in many comparable
calculations, which include diffusion, 
as we were informed by private communication by S.~Degl'Innocenti, S.~Cassisi,
M.~Salaris and I.~Mazzitelli.}. \citen{sgw:00} recently have
investigated in detail the proper use of isochrones to be fitted to
either GC or field halo subdwarf data, when diffusion is included. The
main point to be stressed is that the present surface metallicity of an
individual subdwarf is not the initial one, but is lower by 0.1--0.4
dex (depending on mass and age) due to diffusion. In a GC, however, [Fe/H]
is usually determined from red giants, in which the original surface metallicity
has been restored by convection. Here, an evolutionary track with this
initial metallicity and diffusion included would be the correct one to
be used for an individual star. 

Arguments in favor of diffusion acting close to how it is calculated
are the solar model (\cite{rvc:96}; \cite{gd:97}) and the temperatures
of main-sequence subgiants with HIPPARCOS-distances, which, according
to \citen{mb:99} and \citen{sgw:00} can be explained by the fact that
diffusion leads to lower temperatures (see Fig.~\ref{f:3}). Arguments
against the full action of sedimentation (rather, arguments in favour
of an additional mixing process counteracting diffusion) are the
remaining discrepancy between solar models and the seismic Sun just
below the convective envelope (\cite{rvc:96}) and the presence of
$^7{\rm Li}$ in old metal-poor stars (\cite{vc:98}). In addition, we
remark that \citen{mb:99} tried to explain temperature differences of
about 100~K, while in metal-poor low-mass stars the effect of
diffusion might reduce $T_{\rm eff}$ by 200~K or more. This leads to
colours so red that the comparison with the turn-off colour of some
globular clusters would yield negative reddening (e.g.\ M5, for which
models without diffusion result in a reddening of only 0.02~mag).  We
added a few test calculations (for the case $M/M_\odot=0.8$, $Y=0.25$,
$Z=0.001$), in which either convective overshooting (as in
\cite{schw:99}) or an enhanced stellar wind (following \cite{vc:95})
or both was employed to reduce the effect of gravitational
settling. For pure diffusion a surface helium abundance at the end of
the main sequence of $Y_{\rm s} = 10^{-4}$ results; the overshooting
models retain up to $Y_{\rm s} = 0.04$, those with a Reimers mass loss
($\eta=0.4$) $Y_{\rm s} = 0.06$, and those with both effects $Y_{\rm
s} = 0.09$. Note that the TO age of this star is only 9.8~Gyr. For a
cosmological TO age of about 12~Gyr the mass would be higher and the
sedimentation effect smaller due to the larger extend of the
convective envelope (see also \cite{sgw:00}).

Without elaborating further on this discussion, the true effectiveness of diffusion
might lead to main-sequence lifetimes somewhere between the extremes of no and
full diffusion. All arguments brought forward here concern the photospheric
properties of stars; however, the evolutionary speed is determined by the
central evolution (diffusion leading to a faster aging by adding helium to the
core). On the other hand, the processes counteracting diffusion near to the
photosphere could do the same at the center (e.g.\ rotation-induced
mixing). Therefore, the true main-sequence life-time might be in between the two
limiting cases investigated here; the difference between them being of order
1~Gyr (see Tables A1--A24). A similar result was obtained by \citen{cd:99}, who
discussed the effect in the case of globular cluster isochrones.

To conclude, we have presented an extensive grid of metal-poor low-mass stellar
models. The intention is that these data could be used for determining stellar
ages, if global parameters such as mass, luminosity and composition of
individual halo or globular cluster stars are known. The data can also be used
for standard isochrone construction. To facilitate age derivation, we have
presented fitting formulae, which reproduce the evolutionary results with an
accuracy of 5\% or better for the age range of interest ($\approx
10$~Gyr). We consider the uncertainty of the evolutionary ages to be of order
1~Gyr (at cosmic ages) due to systematic uncertainties in the models and
calculations and another 1~Gyr (at most) due to the unknown effectiveness of
diffusion. In this respect, the accuracy of the fitting formulae is within these
principal uncertainties.

Application of our fit formulae to five metal-poor (halo) field stars with
accurately known metallicity and brightness and reasonably
well-determined mass resulted in ages between 9.7 and 13.4 Gyr (except
for one star with a metallicity outside our model grid). Such ages
appear to be in reasonable agreement with recent globular cluster age
determinations (\cite{sw:98}) using similar stellar models. The
uncertainties due to metallicity or distance errors are smaller than
the model uncertainties, but the 5\% mass uncertainty results in an age
error of up to $\pm4$~Gyr. A 1\% accuracy in mass must be achieved to
make this error source comparable to all others. It appears that the
use of the fitting formulae does not introduce an additional error
source of relevance.

We have not discussed the importance of the effective temperatures, which in the
stellar models depend on the convection theory or parameter used. This is
because $T_{\rm eff}$ is a very insensitive discriminator between different
masses; it should therefore not be used to select the mass of the evolutionary
track to be compared. On the other hand, if the stellar mass is known (with some
error), most likely the effective temperature of the corresponding track is
within the error range. Finally, for known mass, errors in the models' effective
temperature do not influence the $t$--$L$--relation. This is different from
isochrone age determinations, where $T_{\rm eff}$ influences the morphology and
therefore the luminosity of the turn-off, as illustrated by \citen{mdc:95} 
by using two different convection theories. However, $T_{\rm eff}$-values for 
stars with determined mass and luminosity will provide independent checks for
the quality of the stellar models. Our test application results in
effective temperatures, which are within the given uncertainty of the
observationally determined ones. This we regard as an encouraging
confirmation of our models.

\begin{acknowledgements}
The continuing support by B.~Paczy\'nski, who stimulated this work, is
gratefully acknowledged.
We are also indebted to S.~Degl'Innocenti and
M.~Salaris for comparing their results with ours, and to
D.~Alexander and F.~Rogers for providing us with their opacity tables
calculated for our particular needs. H.~Ritter very diligently read the
paper and helped to improve it considerably.
\end{acknowledgements}
%

%
\appendix
\section{Tables containing evolutionary data}

The evolutionary properties for {\em all} evolutionary models
calculated. They are given as Tables~A1--A24 (A1--A12 for the
(canonical) case ``C'' without diffusion and A13--A24 for case ``D'',
the one including diffusion in the same form as Tab.~\ref{t:1} (which
is the table for the C-case composition
$(Y,\,Z)=(0.20,\,0.0001)$). The columns relate to:
\begin{enumerate}	
\item age (in $10^9$~years)
\item $\log L/L_\odot$;
\item $\log T_{\rm eff}$;
\item central helium abundance $Y_{\rm c}$ or relative mass of the
hydrogen-exhausted core, $m_{\rm hc}$; the switching between the two
quantities is easily recognizable by the jump from a number close to 1
to one of order 0.2;
\item surface helium abundance $Y_{\rm s}$;
\item $m_{\rm ce}$, the location of the bottom of the convective
envelope in relative mass coordinate; 
$m_{\rm ce}=1.0$ corresponds to a convective
envelope of thickness $< 10^{-4} M_r/M$ or to a completely radiative
envelope. In a few cases, the very last model exhibits already a
convective shell in the helium flash region. In these cases, this column
gives the bottom of this inner convective region and the value listed is
of order 0.2 or smaller;
\item relative mass of the convective core, $m_{\rm cc}$.
\end{enumerate}
\noindent The tables are an extract of the full evolutionary results,
such that sufficiently, but not too many time-steps are provided. They
contain the most important stages, like ZAMS, turn-off and tip of the
RGB. The complete set of data can be obtained from the authors on
request.

{\renewcommand\baselinestretch{1.0}
\begin{table}[b]
\caption{Evolution of models without diffusion (C-case) of composition
$(Y,\,Z)=(0.20,\,0.001)$} 
\protect\label{t:1}
{\tiny
\begin{flushleft}
\begin{tabular}{rrrrrrr}
Age & $\log L/L_\odot$ & $\lg T_{\rm eff}$ &  $Y_{\rm c}$/$m_{\rm
hc}$& $Y_{\rm s}$ & $m_{\rm ce}$ & $m_{\rm cc}$ \\ \noalign{\smallskip}
\hline \noalign{\smallskip}
\noalign {\bf $M = 0.6\,M_\odot$}
\hline \noalign{\smallskip}
     0.000E+00 &  -0.8711 &   3.7071 &   0.2000 &    0.200 &   0.8967 &   0.0026 \\
     3.308E+08 &  -0.8662 &   3.7045 &   0.2052 &    0.200 &   0.8792 &   0.0690 \\
     1.075E+10 &  -0.7727 &   3.7147 &   0.4375 &    0.200 &   0.9081 &   0.0000 \\
     1.975E+10 &  -0.6792 &   3.7250 &   0.6434 &    0.200 &   0.9323 &   0.0000 \\
     2.694E+10 &  -0.5794 &   3.7351 &   0.7968 &    0.200 &   0.9526 &   0.0000 \\
     3.231E+10 &  -0.4789 &   3.7454 &   0.8988 &    0.200 &   0.9627 &   0.0000 \\
     3.634E+10 &  -0.3772 &   3.7541 &   0.9521 &    0.200 &   0.9716 &   0.0000 \\
     3.927E+10 &  -0.2747 &   3.7608 &   0.9751 &    0.200 &   0.9755 &   0.0000 \\
     4.137E+10 &  -0.1724 &   3.7652 &   0.9868 &    0.200 &   0.9790 &   0.0000 \\
     4.298E+10 &  -0.0697 &   3.7668 &   0.2154 &    0.200 &   0.9818 &   0.0000 \\
     4.409E+10 &   0.0315 &   3.7658 &   0.2361 &    0.200 &   0.9790 &   0.0000 \\
     4.481E+10 &   0.1329 &   3.7601 &   0.2517 &    0.200 &   0.9661 &   0.0000 \\
     4.524E+10 &   0.2338 &   3.7507 &   0.2622 &    0.200 &   0.9322 &   0.0000 \\
     4.553E+10 &   0.3345 &   3.7417 &   0.2702 &    0.200 &   0.8640 &   0.0000 \\
     4.575E+10 &   0.4371 &   3.7357 &   0.2792 &    0.200 &   0.7885 &   0.0000 \\
     4.594E+10 &   0.5377 &   3.7322 &   0.2914 &    0.200 &   0.7269 &   0.0000 \\
     4.612E+10 &   0.6377 &   3.7296 &   0.3075 &    0.200 &   0.6821 &   0.0000 \\
     4.626E+10 &   0.7388 &   3.7273 &   0.3255 &    0.201 &   0.6482 &   0.0000 \\
     4.638E+10 &   0.8398 &   3.7250 &   0.3432 &    0.201 &   0.6204 &   0.0000 \\
     4.647E+10 &   0.9406 &   3.7227 &   0.3617 &    0.201 &   0.5982 &   0.0000 \\
     4.655E+10 &   1.0410 &   3.7203 &   0.3794 &    0.202 &   0.5819 &   0.0000 \\
     4.661E+10 &   1.1414 &   3.7178 &   0.3971 &    0.202 &   0.5712 &   0.0000 \\
     4.666E+10 &   1.2415 &   3.7150 &   0.4143 &    0.202 &   0.5658 &   0.0000 \\
     4.670E+10 &   1.3419 &   3.7122 &   0.4312 &    0.202 &   0.5631 &   0.0000 \\
     4.673E+10 &   1.4420 &   3.7093 &   0.4479 &    0.202 &   0.5631 &   0.0000 \\
     4.676E+10 &   1.5422 &   3.7062 &   0.4648 &    0.202 &   0.5657 &   0.0000 \\
     4.678E+10 &   1.6423 &   3.7031 &   0.4821 &    0.202 &   0.5710 &   0.0000 \\
     4.679E+10 &   1.7425 &   3.6998 &   0.4997 &    0.202 &   0.5790 &   0.0000 \\
     4.681E+10 &   1.8426 &   3.6961 &   0.5174 &    0.202 &   0.5883 &   0.0000 \\
     4.682E+10 &   1.9427 &   3.6927 &   0.5359 &    0.202 &   0.5992 &   0.0000 \\
     4.683E+10 &   2.0427 &   3.6893 &   0.5570 &    0.202 &   0.6144 &   0.0000 \\
     4.684E+10 &   2.1428 &   3.6857 &   0.5755 &    0.202 &   0.6282 &   0.0000 \\
     4.684E+10 &   2.2430 &   3.6821 &   0.5948 &    0.202 &   0.6421 &   0.0000 \\
     4.685E+10 &   2.3431 &   3.6784 &   0.6148 &    0.202 &   0.6576 &   0.0000 \\
     4.685E+10 &   2.4432 &   3.6746 &   0.6356 &    0.202 &   0.6745 &   0.0000 \\
     4.686E+10 &   2.5434 &   3.6707 &   0.6574 &    0.202 &   0.6927 &   0.0000 \\
     4.686E+10 &   2.6435 &   3.6667 &   0.6803 &    0.202 &   0.7123 &   0.0000 \\
     4.686E+10 &   2.7437 &   3.6626 &   0.7039 &    0.202 &   0.7334 &   0.0000 \\
     4.686E+10 &   2.8438 &   3.6583 &   0.7280 &    0.202 &   0.7547 &   0.0000 \\
     4.687E+10 &   2.9440 &   3.6539 &   0.7494 &    0.202 &   0.7733 &   0.0000 \\
     4.687E+10 &   3.0441 &   3.6495 &   0.7736 &    0.202 &   0.7950 &   0.0000 \\
     4.687E+10 &   3.1443 &   3.6452 &   0.7984 &    0.202 &   0.8175 &   0.0000 \\
     4.687E+10 &   3.2445 &   3.6411 &   0.8240 &    0.202 &   0.8411 &   0.0000 \\
     4.687E+10 &   3.3210 &   3.6383 &   0.8435 &    0.202 &   0.8592 &   0.0000 \\
\noalign{\medskip} \hline \noalign{\smallskip}
\noalign {\bf $M = 0.7\,M_\odot$}
\hline \noalign{\smallskip}
     0.000E+00 &  -0.5366 &   3.7473 &   0.2003 &    0.200 &   0.9748 &   0.1191 \\
     1.772E+08 &  -0.5766 &   3.7382 &   0.2046 &    0.200 &   0.9621 &   0.0642 \\
     5.700E+09 &  -0.4884 &   3.7485 &   0.3920 &    0.200 &   0.9732 &   0.0000 \\
     1.078E+10 &  -0.3992 &   3.7589 &   0.5948 &    0.200 &   0.9802 &   0.0000 \\
     1.528E+10 &  -0.2991 &   3.7688 &   0.7709 &    0.200 &   0.9866 &   0.0000 \\
     1.875E+10 &  -0.1978 &   3.7763 &   0.8909 &    0.200 &   0.9916 &   0.0000 \\
     2.127E+10 &  -0.0951 &   3.7829 &   0.9469 &    0.200 &   0.9943 &   0.0000 \\
     2.305E+10 &   0.0075 &   3.7874 &   0.9719 &    0.200 &   0.9959 &   0.0000 \\
     2.435E+10 &   0.1081 &   3.7899 &   0.9905 &    0.200 &   0.9969 &   0.0000 \\
     2.535E+10 &   0.2106 &   3.7893 &   0.2108 &    0.200 &   0.9971 &   0.0000 \\
     2.601E+10 &   0.3127 &   3.7825 &   0.2301 &    0.200 &   0.9953 &   0.0000 \\
     2.632E+10 &   0.3858 &   3.7720 &   0.2400 &    0.200 &   0.9877 &   0.0000 \\
     2.648E+10 &   0.4387 &   3.7614 &   0.2451 &    0.200 &   0.9690 &   0.0000 \\
     2.661E+10 &   0.4915 &   3.7511 &   0.2489 &    0.200 &   0.9273 &   0.0000 \\
     2.674E+10 &   0.5734 &   3.7410 &   0.2533 &    0.200 &   0.8408 &   0.0000 \\
     2.687E+10 &   0.6751 &   3.7346 &   0.2599 &    0.200 &   0.7469 &   0.0000 \\
     2.698E+10 &   0.7766 &   3.7307 &   0.2709 &    0.201 &   0.6742 &   0.0000 \\
     2.707E+10 &   0.8767 &   3.7277 &   0.2855 &    0.201 &   0.6166 &   0.0000 \\
     2.715E+10 &   0.9776 &   3.7249 &   0.3015 &    0.202 &   0.5782 &   0.0000 \\
     2.722E+10 &   1.0782 &   3.7222 &   0.3179 &    0.202 &   0.5467 &   0.0000 \\
     2.728E+10 &   1.1789 &   3.7195 &   0.3341 &    0.203 &   0.5281 &   0.0000 \\
     2.732E+10 &   1.2792 &   3.7167 &   0.3500 &    0.203 &   0.5098 &   0.0000 \\
     2.736E+10 &   1.3796 &   3.7138 &   0.3654 &    0.203 &   0.5006 &   0.0000 \\
     2.738E+10 &   1.4800 &   3.7107 &   0.3807 &    0.203 &   0.4976 &   0.0000 \\
     2.741E+10 &   1.5801 &   3.7076 &   0.3960 &    0.203 &   0.4975 &   0.0000 \\
     2.743E+10 &   1.6803 &   3.7044 &   0.4115 &    0.203 &   0.5005 &   0.0000 \\
     2.744E+10 &   1.7804 &   3.7011 &   0.4273 &    0.203 &   0.5035 &   0.0000 \\
     2.745E+10 &   1.8807 &   3.6973 &   0.4435 &    0.203 &   0.5095 &   0.0000 \\
     2.747E+10 &   1.9809 &   3.6938 &   0.4601 &    0.203 &   0.5186 &   0.0000 \\
     2.747E+10 &   2.0811 &   3.6903 &   0.4777 &    0.203 &   0.5293 &   0.0000 \\
     2.748E+10 &   2.1813 &   3.6868 &   0.4992 &    0.203 &   0.5463 &   0.0000 \\
     2.749E+10 &   2.2814 &   3.6831 &   0.5156 &    0.203 &   0.5580 &   0.0000 \\
     2.749E+10 &   2.3814 &   3.6793 &   0.5328 &    0.203 &   0.5713 &   0.0000 \\
     2.750E+10 &   2.4815 &   3.6754 &   0.5509 &    0.203 &   0.5856 &   0.0000 \\
     2.750E+10 &   2.5816 &   3.6714 &   0.5699 &    0.203 &   0.6016 &   0.0000 \\
     2.750E+10 &   2.6816 &   3.6673 &   0.5897 &    0.203 &   0.6177 &   0.0000 \\
     2.751E+10 &   2.7817 &   3.6631 &   0.6100 &    0.203 &   0.6356 &   0.0000 \\
     2.751E+10 &   2.8817 &   3.6586 &   0.6290 &    0.203 &   0.6518 &   0.0000 \\
     2.751E+10 &   2.9818 &   3.6539 &   0.6490 &    0.203 &   0.6693 &   0.0000 \\
     2.751E+10 &   3.0820 &   3.6492 &   0.6697 &    0.203 &   0.6878 &   0.0000 \\
     2.751E+10 &   3.1821 &   3.6444 &   0.6914 &    0.203 &   0.7075 &   0.0000 \\
     2.751E+10 &   3.2823 &   3.6395 &   0.7131 &    0.203 &   0.7275 &   0.0000 \\
     2.751E+10 &   3.3165 &   3.6379 &   0.7206 &    0.203 &   0.7343 &   0.0000 \\
\end{tabular}
\end{flushleft}
}
\end{table}

\begin{table}{\setcounter{table}{0}}
\caption{(contd.)}
{\tiny
\begin{flushleft}
\begin{tabular}{rrrrrrr}

Age & $\log L/L_\odot$ & $\lg T_{\rm eff}$ &  $Y_{\rm c}$/$m_{\rm
hc}$& $Y_{\rm s}$ & $m_{\rm ce}$ & $m_{\rm cc}$ \\  \noalign{\smallskip}
\hline \noalign{\smallskip}
\noalign {\bf $M = 0.8\,M_\odot$}
\hline \noalign{\smallskip}
     0.000E+00 &  -0.2959 &   3.7834 &   0.2000 &    0.200 &   0.9950 &   0.0828 \\
     1.031E+08 &  -0.3076 &   3.7737 &   0.2042 &    0.200 &   0.9908 &   0.0559 \\
     4.127E+09 &  -0.2068 &   3.7838 &   0.4347 &    0.200 &   0.9950 &   0.0000 \\
     7.556E+09 &  -0.1064 &   3.7933 &   0.6599 &    0.200 &   0.9973 &   0.0000 \\
     1.018E+10 &  -0.0056 &   3.8011 &   0.8284 &    0.200 &   0.9986 &   0.0000 \\
     1.206E+10 &   0.0946 &   3.8074 &   0.9159 &    0.200 &   0.9993 &   0.0000 \\
     1.340E+10 &   0.1947 &   3.8122 &   0.9554 &    0.200 &   0.9996 &   0.0000 \\
     1.441E+10 &   0.2950 &   3.8151 &   0.9843 &    0.200 &   0.9997 &   0.0000 \\
     1.517E+10 &   0.3955 &   3.8146 &   0.1932 &    0.200 &   0.9997 &   0.0000 \\
     1.568E+10 &   0.4963 &   3.8061 &   0.2144 &    0.200 &   0.9996 &   0.0000 \\
     1.587E+10 &   0.5485 &   3.7960 &   0.2227 &    0.200 &   0.9993 &   0.0000 \\
     1.597E+10 &   0.5843 &   3.7858 &   0.2272 &    0.200 &   0.9980 &   0.0000 \\
     1.604E+10 &   0.6136 &   3.7756 &   0.2301 &    0.200 &   0.9937 &   0.0000 \\
     1.610E+10 &   0.6408 &   3.7652 &   0.2323 &    0.200 &   0.9819 &   0.0000 \\
     1.615E+10 &   0.6707 &   3.7550 &   0.2341 &    0.200 &   0.9495 &   0.0000 \\
     1.621E+10 &   0.7177 &   3.7450 &   0.2360 &    0.200 &   0.8792 &   0.0000 \\
     1.630E+10 &   0.8185 &   3.7357 &   0.2397 &    0.200 &   0.7525 &   0.0000 \\
     1.637E+10 &   0.9191 &   3.7308 &   0.2466 &    0.201 &   0.6552 &   0.0000 \\
     1.643E+10 &   1.0195 &   3.7273 &   0.2586 &    0.201 &   0.5846 &   0.0000 \\
     1.649E+10 &   1.1200 &   3.7242 &   0.2729 &    0.202 &   0.5344 &   0.0000 \\
     1.653E+10 &   1.2203 &   3.7212 &   0.2875 &    0.203 &   0.4978 &   0.0000 \\
     1.657E+10 &   1.3209 &   3.7182 &   0.3019 &    0.204 &   0.4736 &   0.0000 \\
     1.660E+10 &   1.4209 &   3.7152 &   0.3161 &    0.204 &   0.4560 &   0.0000 \\
     1.663E+10 &   1.5212 &   3.7121 &   0.3302 &    0.205 &   0.4472 &   0.0000 \\
     1.665E+10 &   1.6213 &   3.7089 &   0.3443 &    0.205 &   0.4443 &   0.0000 \\
     1.667E+10 &   1.7215 &   3.7057 &   0.3585 &    0.205 &   0.4442 &   0.0000 \\
     1.668E+10 &   1.8216 &   3.7023 &   0.3730 &    0.205 &   0.4471 &   0.0000 \\
     1.669E+10 &   1.9217 &   3.6984 &   0.3878 &    0.205 &   0.4500 &   0.0000 \\
     1.670E+10 &   2.0220 &   3.6949 &   0.4030 &    0.205 &   0.4572 &   0.0000 \\
     1.671E+10 &   2.1222 &   3.6913 &   0.4188 &    0.205 &   0.4658 &   0.0000 \\
     1.672E+10 &   2.2223 &   3.6879 &   0.4426 &    0.205 &   0.4868 &   0.0000 \\
     1.673E+10 &   2.3224 &   3.6841 &   0.4569 &    0.205 &   0.4951 &   0.0000 \\
     1.673E+10 &   2.4226 &   3.6802 &   0.4722 &    0.205 &   0.5064 &   0.0000 \\
     1.673E+10 &   2.5227 &   3.6763 &   0.4883 &    0.205 &   0.5193 &   0.0000 \\
     1.674E+10 &   2.6228 &   3.6723 &   0.5052 &    0.205 &   0.5323 &   0.0000 \\
     1.674E+10 &   2.7229 &   3.6681 &   0.5226 &    0.205 &   0.5474 &   0.0000 \\
     1.674E+10 &   2.8229 &   3.6638 &   0.5403 &    0.205 &   0.5626 &   0.0000 \\
     1.674E+10 &   2.9230 &   3.6592 &   0.5569 &    0.205 &   0.5768 &   0.0000 \\
     1.674E+10 &   3.0231 &   3.6545 &   0.5747 &    0.205 &   0.5923 &   0.0000 \\
     1.675E+10 &   3.1231 &   3.6496 &   0.5932 &    0.205 &   0.6088 &   0.0000 \\
     1.675E+10 &   3.2233 &   3.6446 &   0.6121 &    0.205 &   0.6258 &   0.0000 \\
     1.675E+10 &   3.3113 &   3.6402 &   0.6287 &    0.205 &   0.6409 &   0.0000 \\
\noalign{\medskip} \hline \noalign{\smallskip}
\noalign {\bf $M = 0.9\,M_\odot$}
\hline \noalign{\smallskip}
     0.000E+00 &  -0.0712 &   3.8094 &   0.2000 &    0.200 &   0.9992 &   0.1091 \\
     1.088E+08 &  -0.0713 &   3.8036 &   0.2065 &    0.200 &   0.9988 &   0.0543 \\
     2.919E+09 &   0.0295 &   3.8138 &   0.4335 &    0.200 &   0.9995 &   0.0000 \\
     5.271E+09 &   0.1304 &   3.8227 &   0.6745 &    0.200 &   0.9997 &   0.0000 \\
     7.046E+09 &   0.2323 &   3.8304 &   0.8425 &    0.200 &   0.9998 &   0.0000 \\
     8.314E+09 &   0.3352 &   3.8372 &   0.9209 &    0.200 &   0.9998 &   0.0000 \\
     9.224E+09 &   0.4366 &   3.8425 &   0.9646 &    0.200 &   1.0000 &   0.0000 \\
     9.929E+09 &   0.5372 &   3.8448 &   0.1724 &    0.200 &   1.0000 &   0.0000 \\
     1.042E+10 &   0.6380 &   3.8379 &   0.1979 &    0.200 &   1.0000 &   0.0000 \\
     1.060E+10 &   0.6895 &   3.8272 &   0.2080 &    0.200 &   0.9998 &   0.0000 \\
     1.069E+10 &   0.7230 &   3.8168 &   0.2132 &    0.200 &   0.9998 &   0.0000 \\
     1.075E+10 &   0.7494 &   3.8063 &   0.2165 &    0.200 &   0.9998 &   0.0000 \\
     1.080E+10 &   0.7726 &   3.7953 &   0.2188 &    0.200 &   0.9995 &   0.0000 \\
     1.083E+10 &   0.7916 &   3.7850 &   0.2204 &    0.200 &   0.9986 &   0.0000 \\
     1.086E+10 &   0.8092 &   3.7743 &   0.2217 &    0.200 &   0.9950 &   0.0000 \\
     1.089E+10 &   0.8250 &   3.7639 &   0.2227 &    0.200 &   0.9835 &   0.0000 \\
     1.092E+10 &   0.8443 &   3.7536 &   0.2236 &    0.200 &   0.9495 &   0.0000 \\
     1.095E+10 &   0.8840 &   3.7434 &   0.2246 &    0.200 &   0.8666 &   0.0000 \\
     1.101E+10 &   0.9842 &   3.7340 &   0.2266 &    0.200 &   0.7150 &   0.0000 \\
     1.106E+10 &   1.0845 &   3.7292 &   0.2314 &    0.201 &   0.6103 &   0.0000 \\
     1.110E+10 &   1.1855 &   3.7255 &   0.2422 &    0.202 &   0.5345 &   0.0000 \\
     1.114E+10 &   1.2863 &   3.7222 &   0.2550 &    0.204 &   0.4797 &   0.0000 \\
     1.117E+10 &   1.3867 &   3.7190 &   0.2680 &    0.205 &   0.4444 &   0.0000 \\
     1.119E+10 &   1.4873 &   3.7158 &   0.2811 &    0.206 &   0.4217 &   0.0000 \\
     1.121E+10 &   1.5878 &   3.7126 &   0.2941 &    0.206 &   0.4106 &   0.0000 \\
     1.123E+10 &   1.6878 &   3.7094 &   0.3071 &    0.207 &   0.4023 &   0.0000 \\
     1.125E+10 &   1.7881 &   3.7060 &   0.3203 &    0.207 &   0.3995 &   0.0000 \\
     1.126E+10 &   1.8881 &   3.7025 &   0.3339 &    0.207 &   0.4022 &   0.0000 \\
     1.127E+10 &   1.9884 &   3.6986 &   0.3477 &    0.207 &   0.4050 &   0.0000 \\
     1.128E+10 &   2.0885 &   3.6950 &   0.3619 &    0.207 &   0.4104 &   0.0000 \\
     1.128E+10 &   2.1887 &   3.6914 &   0.3766 &    0.207 &   0.4187 &   0.0000 \\
     1.129E+10 &   2.2889 &   3.6879 &   0.4020 &    0.207 &   0.4413 &   0.0000 \\
     1.130E+10 &   2.3889 &   3.6840 &   0.4151 &    0.207 &   0.4491 &   0.0000 \\
     1.130E+10 &   2.4891 &   3.6801 &   0.4291 &    0.207 &   0.4592 &   0.0000 \\
     1.130E+10 &   2.5892 &   3.6761 &   0.4437 &    0.207 &   0.4702 &   0.0000 \\
     1.131E+10 &   2.6893 &   3.6720 &   0.4589 &    0.207 &   0.4830 &   0.0000 \\
     1.131E+10 &   2.7894 &   3.6678 &   0.4744 &    0.207 &   0.4958 &   0.0000 \\
     1.131E+10 &   2.8895 &   3.6633 &   0.4896 &    0.207 &   0.5086 &   0.0000 \\
     1.131E+10 &   2.9895 &   3.6587 &   0.5050 &    0.207 &   0.5219 &   0.0000 \\
     1.131E+10 &   3.0898 &   3.6539 &   0.5214 &    0.207 &   0.5361 &   0.0000 \\
     1.131E+10 &   3.1901 &   3.6489 &   0.5380 &    0.207 &   0.5509 &   0.0000 \\
     1.132E+10 &   3.2903 &   3.6438 &   0.5548 &    0.207 &   0.5661 &   0.0000 \\
     1.132E+10 &   3.3799 &   3.6391 &   0.5705 &    0.207 &   0.1554 &   0.0000 \\
\end{tabular}
\end{flushleft}
}
\end{table}

\begin{table}{\setcounter{table}{0}}
\caption{(contd.)}
{\tiny
\begin{flushleft}
\begin{tabular}{rrrrrrr}

Age & $\log L/L_\odot$ & $\lg T_{\rm eff}$ &  $Y_{\rm c}$/$m_{\rm
hc}$& $Y_{\rm s}$ & $m_{\rm ce}$ & $m_{\rm cc}$ \\  \noalign{\smallskip}
\hline \noalign{\smallskip}
\noalign {\bf $M = 1.0\,M_\odot$}
\hline \noalign{\smallskip}
     0.000E+00 &   0.1257 &   3.8330 &   0.2000 &    0.200 &   0.9998 &   0.1296 \\
     1.088E+08 &   0.1379 &   3.8327 &   0.2092 &    0.200 &   0.9998 &   0.0575 \\
     1.907E+09 &   0.2274 &   3.8429 &   0.4011 &    0.200 &   0.9998 &   0.0014 \\
     3.647E+09 &   0.3305 &   3.8531 &   0.6710 &    0.200 &   1.0000 &   0.0000 \\
     4.906E+09 &   0.4331 &   3.8633 &   0.8411 &    0.200 &   1.0000 &   0.0000 \\
     5.748E+09 &   0.5290 &   3.8735 &   0.9178 &    0.200 &   1.0000 &   0.0000 \\
     6.449E+09 &   0.6315 &   3.8819 &   0.9810 &    0.200 &   1.0000 &   0.0000 \\
     6.977E+09 &   0.7343 &   3.8824 &   0.1740 &    0.200 &   1.0000 &   0.0000 \\
     7.246E+09 &   0.8113 &   3.8720 &   0.1925 &    0.200 &   1.0000 &   0.0000 \\
     7.337E+09 &   0.8463 &   3.8613 &   0.1989 &    0.200 &   1.0000 &   0.0000 \\
     7.391E+09 &   0.8712 &   3.8500 &   0.2027 &    0.200 &   1.0000 &   0.0000 \\
     7.423E+09 &   0.8885 &   3.8398 &   0.2050 &    0.200 &   1.0000 &   0.0000 \\
     7.449E+09 &   0.9043 &   3.8289 &   0.2067 &    0.200 &   1.0000 &   0.0000 \\
     7.472E+09 &   0.9195 &   3.8176 &   0.2082 &    0.200 &   0.9998 &   0.0000 \\
     7.491E+09 &   0.9327 &   3.8075 &   0.2093 &    0.200 &   0.9998 &   0.0000 \\
     7.509E+09 &   0.9461 &   3.7967 &   0.2103 &    0.200 &   0.9997 &   0.0000 \\
     7.525E+09 &   0.9579 &   3.7862 &   0.2110 &    0.200 &   0.9993 &   0.0000 \\
     7.539E+09 &   0.9684 &   3.7756 &   0.2117 &    0.200 &   0.9973 &   0.0000 \\
     7.555E+09 &   0.9777 &   3.7649 &   0.2123 &    0.200 &   0.9892 &   0.0000 \\
     7.571E+09 &   0.9881 &   3.7542 &   0.2128 &    0.200 &   0.9597 &   0.0000 \\
     7.593E+09 &   1.0142 &   3.7439 &   0.2133 &    0.200 &   0.8822 &   0.0000 \\
     7.631E+09 &   1.1026 &   3.7338 &   0.2144 &    0.200 &   0.7150 &   0.0000 \\
     7.665E+09 &   1.2035 &   3.7285 &   0.2170 &    0.201 &   0.5910 &   0.0000 \\
     7.696E+09 &   1.3045 &   3.7247 &   0.2257 &    0.203 &   0.5040 &   0.0000 \\
     7.722E+09 &   1.4049 &   3.7212 &   0.2370 &    0.205 &   0.4503 &   0.0000 \\
     7.745E+09 &   1.5053 &   3.7179 &   0.2488 &    0.206 &   0.4107 &   0.0000 \\
     7.764E+09 &   1.6059 &   3.7146 &   0.2607 &    0.208 &   0.3867 &   0.0000 \\
     7.779E+09 &   1.7063 &   3.7113 &   0.2726 &    0.208 &   0.3746 &   0.0000 \\
     7.793E+09 &   1.8067 &   3.7078 &   0.2849 &    0.209 &   0.3665 &   0.0000 \\
     7.804E+09 &   1.9067 &   3.7043 &   0.2974 &    0.209 &   0.3665 &   0.0000 \\
     7.813E+09 &   2.0069 &   3.7007 &   0.3101 &    0.209 &   0.3678 &   0.0000 \\
     7.821E+09 &   2.1070 &   3.6968 &   0.3232 &    0.209 &   0.3718 &   0.0000 \\
     7.827E+09 &   2.2073 &   3.6931 &   0.3365 &    0.209 &   0.3771 &   0.0000 \\
     7.833E+09 &   2.3074 &   3.6893 &   0.3514 &    0.209 &   0.3855 &   0.0000 \\
     7.842E+09 &   2.4076 &   3.6857 &   0.3762 &    0.209 &   0.4083 &   0.0000 \\
     7.846E+09 &   2.5077 &   3.6818 &   0.3887 &    0.209 &   0.4166 &   0.0000 \\
     7.849E+09 &   2.6078 &   3.6777 &   0.4018 &    0.209 &   0.4263 &   0.0000 \\
     7.851E+09 &   2.7079 &   3.6736 &   0.4154 &    0.209 &   0.4377 &   0.0000 \\
     7.853E+09 &   2.8080 &   3.6694 &   0.4293 &    0.209 &   0.4487 &   0.0000 \\
     7.855E+09 &   2.9081 &   3.6650 &   0.4429 &    0.209 &   0.4599 &   0.0000 \\
     7.856E+09 &   3.0082 &   3.6603 &   0.4569 &    0.209 &   0.4720 &   0.0000 \\
     7.858E+09 &   3.1083 &   3.6556 &   0.4718 &    0.209 &   0.4850 &   0.0000 \\
     7.859E+09 &   3.2084 &   3.6506 &   0.4868 &    0.209 &   0.4982 &   0.0000 \\
     7.859E+09 &   3.2948 &   3.6462 &   0.4996 &    0.209 &   0.5097 &   0.0000 \\
\noalign{\medskip} \hline \noalign{\smallskip}
\noalign {\bf $M = 1.1\,M_\odot$}
\hline \noalign{\smallskip}
     0.000E+00 &   0.3019 &   3.8562 &   0.2000 &    0.200 &   1.0000 &   0.1466 \\
     1.317E+08 &   0.3258 &   3.8631 &   0.2145 &    0.200 &   1.0000 &   0.0673 \\
     1.185E+09 &   0.3939 &   3.8737 &   0.3469 &    0.200 &   1.0000 &   0.0294 \\
     2.105E+09 &   0.4622 &   3.8838 &   0.5187 &    0.200 &   1.0000 &   0.0020 \\
     2.973E+09 &   0.5409 &   3.8940 &   0.7225 &    0.200 &   1.0000 &   0.0000 \\
     3.643E+09 &   0.6205 &   3.9041 &   0.8441 &    0.200 &   1.0000 &   0.0000 \\
     4.281E+09 &   0.7177 &   3.9144 &   0.9355 &    0.200 &   1.0000 &   0.0000 \\
     4.810E+09 &   0.8183 &   3.9192 &   0.1482 &    0.200 &   1.0000 &   0.0000 \\
     5.164E+09 &   0.9207 &   3.9136 &   0.1771 &    0.200 &   1.0000 &   0.0000 \\
     5.280E+09 &   0.9698 &   3.9027 &   0.1873 &    0.200 &   1.0000 &   0.0000 \\
     5.336E+09 &   0.9995 &   3.8912 &   0.1921 &    0.200 &   1.0000 &   0.0000 \\
     5.368E+09 &   1.0206 &   3.8795 &   0.1949 &    0.200 &   1.0000 &   0.0000 \\
     5.389E+09 &   1.0355 &   3.8687 &   0.1966 &    0.200 &   1.0000 &   0.0000 \\
     5.405E+09 &   1.0487 &   3.8570 &   0.1979 &    0.200 &   1.0000 &   0.0000 \\
     5.416E+09 &   1.0587 &   3.8461 &   0.1987 &    0.200 &   1.0000 &   0.0000 \\
     5.426E+09 &   1.0679 &   3.8348 &   0.1994 &    0.200 &   1.0000 &   0.0000 \\
     5.434E+09 &   1.0753 &   3.8247 &   0.2000 &    0.200 &   1.0000 &   0.0000 \\
     5.442E+09 &   1.0833 &   3.8136 &   0.2005 &    0.200 &   0.9998 &   0.0000 \\
     5.451E+09 &   1.0920 &   3.8020 &   0.2010 &    0.200 &   0.9998 &   0.0000 \\
     5.459E+09 &   1.1000 &   3.7908 &   0.2015 &    0.200 &   0.9997 &   0.0000 \\
     5.468E+09 &   1.1069 &   3.7799 &   0.2019 &    0.200 &   0.9990 &   0.0000 \\
     5.477E+09 &   1.1118 &   3.7695 &   0.2023 &    0.200 &   0.9956 &   0.0000 \\
     5.486E+09 &   1.1151 &   3.7591 &   0.2026 &    0.200 &   0.9827 &   0.0000 \\
     5.498E+09 &   1.1217 &   3.7489 &   0.2029 &    0.200 &   0.9386 &   0.0000 \\
     5.515E+09 &   1.1560 &   3.7388 &   0.2033 &    0.200 &   0.8233 &   0.0000 \\
     5.545E+09 &   1.2567 &   3.7305 &   0.2042 &    0.201 &   0.6429 &   0.0000 \\
     5.569E+09 &   1.3575 &   3.7259 &   0.2078 &    0.202 &   0.5284 &   0.0000 \\
     5.590E+09 &   1.4583 &   3.7222 &   0.2169 &    0.204 &   0.4504 &   0.0000 \\
     5.609E+09 &   1.5587 &   3.7187 &   0.2273 &    0.207 &   0.3997 &   0.0000 \\
     5.625E+09 &   1.6593 &   3.7153 &   0.2382 &    0.208 &   0.3720 &   0.0000 \\
     5.638E+09 &   1.7597 &   3.7118 &   0.2493 &    0.210 &   0.3508 &   0.0000 \\
     5.650E+09 &   1.8598 &   3.7084 &   0.2607 &    0.210 &   0.3403 &   0.0000 \\
     5.659E+09 &   1.9598 &   3.7048 &   0.2724 &    0.211 &   0.3377 &   0.0000 \\
     5.667E+09 &   2.0600 &   3.7011 &   0.2842 &    0.211 &   0.3377 &   0.0000 \\
     5.674E+09 &   2.1602 &   3.6971 &   0.2963 &    0.211 &   0.3415 &   0.0000 \\
     5.680E+09 &   2.2604 &   3.6934 &   0.3086 &    0.211 &   0.3467 &   0.0000 \\
     5.685E+09 &   2.3605 &   3.6895 &   0.3225 &    0.211 &   0.3532 &   0.0000 \\
     5.694E+09 &   2.4607 &   3.6859 &   0.3484 &    0.211 &   0.3770 &   0.0000 \\
     5.697E+09 &   2.5608 &   3.6820 &   0.3599 &    0.211 &   0.3849 &   0.0000 \\
     5.700E+09 &   2.6609 &   3.6779 &   0.3718 &    0.211 &   0.3938 &   0.0000 \\
     5.702E+09 &   2.7611 &   3.6737 &   0.3842 &    0.211 &   0.4035 &   0.0000 \\
     5.704E+09 &   2.8612 &   3.6694 &   0.3972 &    0.211 &   0.4145 &   0.0000 \\
     5.705E+09 &   2.9614 &   3.6650 &   0.4093 &    0.211 &   0.4245 &   0.0000 \\
     5.707E+09 &   3.0618 &   3.6603 &   0.4225 &    0.211 &   0.4359 &   0.0000 \\
     5.708E+09 &   3.1620 &   3.6554 &   0.4362 &    0.211 &   0.4476 &   0.0000 \\
     5.709E+09 &   3.2623 &   3.6504 &   0.4496 &    0.211 &   0.4595 &   0.0000 \\
     5.709E+09 &   3.2886 &   3.6491 &   0.4532 &    0.211 &   0.4628 &   0.0000 \\
\end{tabular}
\end{flushleft}
}
\end{table}

\begin{table}{\setcounter{table}{0}}
\caption{(contd.)}
{\tiny
\begin{flushleft}
\begin{tabular}{rrrrrrr}

Age & $\log L/L_\odot$ & $\lg T_{\rm eff}$ &  $Y_{\rm c}$/$m_{\rm
hc}$& $Y_{\rm s}$ & $m_{\rm ce}$ & $m_{\rm cc}$ \\  \noalign{\smallskip}
\hline \noalign{\smallskip}
\noalign {\bf $M = 1.2\,M_\odot$}
\hline \noalign{\smallskip}
     0.000E+00 &   0.4853 &   3.8990 &   0.2000 &    0.200 &   1.0000 &   0.1497 \\
     1.089E+09 &   0.5639 &   3.9094 &   0.3474 &    0.200 &   1.0000 &   0.0655 \\
     1.773E+09 &   0.6281 &   3.9195 &   0.4827 &    0.200 &   1.0000 &   0.0394 \\
     2.358E+09 &   0.6987 &   3.9298 &   0.6531 &    0.200 &   1.0000 &   0.0000 \\
     2.895E+09 &   0.7807 &   3.9400 &   0.8283 &    0.200 &   1.0000 &   0.0000 \\
     3.427E+09 &   0.8812 &   3.9485 &   0.9741 &    0.200 &   1.0000 &   0.0000 \\
     3.636E+09 &   0.9312 &   3.9505 &   0.1344 &    0.200 &   1.0000 &   0.0000 \\
     3.944E+09 &   1.0335 &   3.9467 &   0.1649 &    0.200 &   1.0000 &   0.0000 \\
     4.058E+09 &   1.0895 &   3.9361 &   0.1770 &    0.200 &   1.0000 &   0.0000 \\
     4.104E+09 &   1.1185 &   3.9258 &   0.1819 &    0.200 &   1.0000 &   0.0000 \\
     4.132E+09 &   1.1396 &   3.9149 &   0.1848 &    0.200 &   1.0000 &   0.0000 \\
     4.152E+09 &   1.1567 &   3.9032 &   0.1868 &    0.200 &   1.0000 &   0.0000 \\
     4.165E+09 &   1.1697 &   3.8918 &   0.1881 &    0.200 &   1.0000 &   0.0000 \\
     4.175E+09 &   1.1803 &   3.8805 &   0.1890 &    0.200 &   1.0000 &   0.0000 \\
     4.183E+09 &   1.1891 &   3.8693 &   0.1897 &    0.200 &   1.0000 &   0.0000 \\
     4.189E+09 &   1.1968 &   3.8579 &   0.1902 &    0.200 &   1.0000 &   0.0000 \\
     4.195E+09 &   1.2035 &   3.8461 &   0.1906 &    0.200 &   1.0000 &   0.0000 \\
     4.199E+09 &   1.2086 &   3.8359 &   0.1909 &    0.200 &   1.0000 &   0.0000 \\
     4.203E+09 &   1.2135 &   3.8252 &   0.1912 &    0.200 &   1.0000 &   0.0000 \\
     4.206E+09 &   1.2179 &   3.8149 &   0.1915 &    0.200 &   1.0000 &   0.0000 \\
     4.211E+09 &   1.2226 &   3.8038 &   0.1917 &    0.200 &   0.9998 &   0.0000 \\
     4.216E+09 &   1.2274 &   3.7923 &   0.1920 &    0.200 &   0.9997 &   0.0000 \\
     4.221E+09 &   1.2310 &   3.7817 &   0.1923 &    0.200 &   0.9994 &   0.0000 \\
     4.226E+09 &   1.2328 &   3.7714 &   0.1925 &    0.200 &   0.9977 &   0.0000 \\
     4.232E+09 &   1.2318 &   3.7608 &   0.1928 &    0.200 &   0.9896 &   0.0000 \\
     4.240E+09 &   1.2300 &   3.7507 &   0.1930 &    0.200 &   0.9597 &   0.0000 \\
     4.251E+09 &   1.2446 &   3.7405 &   0.1933 &    0.200 &   0.8600 &   0.0000 \\
     4.275E+09 &   1.3426 &   3.7304 &   0.1939 &    0.201 &   0.6429 &   0.0000 \\
     4.294E+09 &   1.4435 &   3.7256 &   0.1966 &    0.202 &   0.5161 &   0.0000 \\
     4.310E+09 &   1.5439 &   3.7217 &   0.2044 &    0.205 &   0.4332 &   0.0000 \\
     4.325E+09 &   1.6441 &   3.7181 &   0.2138 &    0.207 &   0.3774 &   0.0000 \\
     4.338E+09 &   1.7442 &   3.7147 &   0.2238 &    0.210 &   0.3456 &   0.0000 \\
     4.349E+09 &   1.8446 &   3.7111 &   0.2343 &    0.211 &   0.3277 &   0.0000 \\
     4.359E+09 &   1.9448 &   3.7075 &   0.2450 &    0.212 &   0.3176 &   0.0000 \\
     4.367E+09 &   2.0449 &   3.7039 &   0.2560 &    0.212 &   0.3151 &   0.0000 \\
     4.373E+09 &   2.1451 &   3.7001 &   0.2670 &    0.212 &   0.3151 &   0.0000 \\
     4.379E+09 &   2.2452 &   3.6961 &   0.2782 &    0.212 &   0.3175 &   0.0000 \\
     4.383E+09 &   2.3454 &   3.6923 &   0.2900 &    0.212 &   0.3224 &   0.0000 \\
     4.393E+09 &   2.4456 &   3.6887 &   0.3186 &    0.212 &   0.3479 &   0.0000 \\
     4.396E+09 &   2.5457 &   3.6847 &   0.3287 &    0.212 &   0.3538 &   0.0000 \\
     4.399E+09 &   2.6459 &   3.6806 &   0.3394 &    0.212 &   0.3610 &   0.0000 \\
     4.401E+09 &   2.7460 &   3.6765 &   0.3507 &    0.212 &   0.3696 &   0.0000 \\
     4.403E+09 &   2.8461 &   3.6722 &   0.3624 &    0.212 &   0.3790 &   0.0000 \\
     4.405E+09 &   2.9462 &   3.6678 &   0.3737 &    0.212 &   0.3882 &   0.0000 \\
     4.406E+09 &   3.0462 &   3.6633 &   0.3857 &    0.212 &   0.3985 &   0.0000 \\
     4.407E+09 &   3.1464 &   3.6585 &   0.3979 &    0.212 &   0.4089 &   0.0000 \\
     4.408E+09 &   3.2473 &   3.6535 &   0.4103 &    0.212 &   0.4197 &   0.0000 \\
     4.409E+09 &   3.3395 &   3.6489 &   0.4225 &    0.212 &   0.0978 &   0.0000 \\
\noalign{\medskip} \hline \noalign{\smallskip}
\noalign {\bf $M = 1.3\,M_\odot$}
\hline \noalign{\smallskip}
     0.000E+00 &   0.6437 &   3.9350 &   0.2006 &    0.200 &   1.0000 &   0.0856 \\
     1.017E+08 &   0.6374 &   3.9278 &   0.2147 &    0.200 &   1.0000 &   0.1170 \\
     8.388E+08 &   0.7009 &   3.9384 &   0.3248 &    0.200 &   1.0000 &   0.0938 \\
     1.474E+09 &   0.7684 &   3.9490 &   0.4575 &    0.200 &   1.0000 &   0.0689 \\
     1.953E+09 &   0.8353 &   3.9594 &   0.5987 &    0.200 &   1.0000 &   0.0324 \\
     2.339E+09 &   0.9086 &   3.9695 &   0.7877 &    0.200 &   1.0000 &   0.0000 \\
     2.782E+09 &   1.0098 &   3.9769 &   0.9912 &    0.200 &   1.0000 &   0.0000 \\
     3.076E+09 &   1.1108 &   3.9778 &   0.1453 &    0.200 &   1.0000 &   0.0000 \\
     3.234E+09 &   1.1924 &   3.9673 &   0.1654 &    0.200 &   1.0000 &   0.0000 \\
     3.280E+09 &   1.2257 &   3.9569 &   0.1714 &    0.200 &   1.0000 &   0.0000 \\
     3.306E+09 &   1.2479 &   3.9463 &   0.1748 &    0.200 &   1.0000 &   0.0000 \\
     3.323E+09 &   1.2652 &   3.9351 &   0.1769 &    0.200 &   1.0000 &   0.0000 \\
     3.334E+09 &   1.2784 &   3.9240 &   0.1782 &    0.200 &   1.0000 &   0.0000 \\
     3.342E+09 &   1.2883 &   3.9139 &   0.1792 &    0.200 &   1.0000 &   0.0000 \\
     3.349E+09 &   1.2973 &   3.9028 &   0.1799 &    0.200 &   1.0000 &   0.0000 \\
     3.355E+09 &   1.3055 &   3.8909 &   0.1805 &    0.200 &   1.0000 &   0.0000 \\
     3.360E+09 &   1.3124 &   3.8789 &   0.1810 &    0.200 &   1.0000 &   0.0000 \\
     3.363E+09 &   1.3179 &   3.8677 &   0.1813 &    0.200 &   1.0000 &   0.0000 \\
     3.367E+09 &   1.3227 &   3.8562 &   0.1816 &    0.200 &   1.0000 &   0.0000 \\
     3.370E+09 &   1.3267 &   3.8451 &   0.1818 &    0.200 &   1.0000 &   0.0000 \\
     3.372E+09 &   1.3301 &   3.8334 &   0.1820 &    0.200 &   1.0000 &   0.0000 \\
     3.375E+09 &   1.3329 &   3.8228 &   0.1822 &    0.200 &   1.0000 &   0.0000 \\
     3.377E+09 &   1.3355 &   3.8113 &   0.1824 &    0.200 &   1.0000 &   0.0000 \\
     3.380E+09 &   1.3376 &   3.8006 &   0.1825 &    0.200 &   0.9998 &   0.0000 \\
     3.383E+09 &   1.3394 &   3.7895 &   0.1827 &    0.200 &   0.9998 &   0.0000 \\
     3.386E+09 &   1.3397 &   3.7787 &   0.1829 &    0.200 &   0.9994 &   0.0000 \\
     3.390E+09 &   1.3374 &   3.7682 &   0.1831 &    0.200 &   0.9975 &   0.0000 \\
     3.395E+09 &   1.3315 &   3.7578 &   0.1832 &    0.200 &   0.9877 &   0.0000 \\
     3.400E+09 &   1.3252 &   3.7477 &   0.1834 &    0.200 &   0.9495 &   0.0000 \\
     3.410E+09 &   1.3446 &   3.7376 &   0.1837 &    0.200 &   0.8281 &   0.0000 \\
     3.429E+09 &   1.4458 &   3.7288 &   0.1845 &    0.201 &   0.6103 &   0.0000 \\
     3.444E+09 &   1.5468 &   3.7242 &   0.1885 &    0.203 &   0.4859 &   0.0000 \\
     3.457E+09 &   1.6471 &   3.7203 &   0.1962 &    0.206 &   0.4053 &   0.0000 \\
     3.469E+09 &   1.7476 &   3.7167 &   0.2051 &    0.209 &   0.3509 &   0.0000 \\
     3.479E+09 &   1.8478 &   3.7131 &   0.2146 &    0.211 &   0.3201 &   0.0000 \\
     3.488E+09 &   1.9478 &   3.7095 &   0.2245 &    0.212 &   0.3029 &   0.0000 \\
     3.496E+09 &   2.0481 &   3.7058 &   0.2346 &    0.213 &   0.2957 &   0.0000 \\
     3.502E+09 &   2.1482 &   3.7021 &   0.2447 &    0.213 &   0.2945 &   0.0000 \\
     3.508E+09 &   2.2485 &   3.6980 &   0.2552 &    0.213 &   0.2957 &   0.0000 \\
     3.512E+09 &   2.3487 &   3.6942 &   0.2660 &    0.213 &   0.2992 &   0.0000 \\
     3.516E+09 &   2.4489 &   3.6902 &   0.2776 &    0.213 &   0.3040 &   0.0000 \\
     3.525E+09 &   2.5490 &   3.6866 &   0.3043 &    0.213 &   0.3287 &   0.0000 \\
     3.527E+09 &   2.6492 &   3.6825 &   0.3140 &    0.213 &   0.3350 &   0.0000 \\
     3.529E+09 &   2.7492 &   3.6783 &   0.3243 &    0.213 &   0.3426 &   0.0000 \\
     3.531E+09 &   2.8494 &   3.6741 &   0.3352 &    0.213 &   0.3511 &   0.0000 \\
     3.533E+09 &   2.9495 &   3.6698 &   0.3457 &    0.213 &   0.3593 &   0.0000 \\
     3.534E+09 &   3.0498 &   3.6652 &   0.3567 &    0.213 &   0.3686 &   0.0000 \\
     3.535E+09 &   3.1498 &   3.6605 &   0.3680 &    0.213 &   0.3783 &   0.0000 \\
     3.536E+09 &   3.2499 &   3.6556 &   0.3793 &    0.213 &   0.3881 &   0.0000 \\
     3.537E+09 &   3.3210 &   3.6521 &   0.3881 &    0.213 &   0.0781 &   0.0000 \\
\end{tabular}
\end{flushleft}
}
\end{table}
}

\end{document}